\documentclass{article} %
\usepackage{iclr2027_conference,times}
\usepackage{xspace}

\usepackage{amsmath,amsfonts,bm}

\def\eqref#1{equation~\ref{#1}}

\def\1{\bm{1}}

\DeclareMathAlphabet{\mathsfit}{\encodingdefault}{\sfdefault}{m}{sl}
\SetMathAlphabet{\mathsfit}{bold}{\encodingdefault}{\sfdefault}{bx}{n}

\iclrfinalcopy %

\usepackage[utf8]{inputenc} %
\usepackage[T1]{fontenc}    %
\usepackage{colortbl}
\usepackage{hyperref}       %
\usepackage{url}            %
\usepackage{booktabs}       %
\usepackage{amsfonts}       %
\usepackage{nicefrac}       %
\usepackage{microtype}      %
\usepackage{graphicx}
\usepackage{amsmath}
\usepackage{multirow}
\usepackage{subcaption}
\usepackage{array}
\newsavebox{\iclrhiddencolbox}
\newcolumntype{H}{>{\begin{lrbox}{\iclrhiddencolbox}}p{0pt}<{\end{lrbox}}@{}}
\usepackage{wrapfig}
\usepackage{tcolorbox}
\usepackage{caption}  %

\newcommand{\myparagraph}[1]{\noindent\textbf{#1}\xspace}

\newcommand{\omnigen}{\textsc{OmniGen2}\xspace}
\newcommand{\bagel}{\textsc{BAGEL}\xspace}

\definecolor{revblue}{RGB}{0,90,190}

\definecolor{revred}{RGB}{190,0,0}

\title{The Poisoned Conversation: Privacy-Leaking Watermarks in Unified Multimodal Models}

\author{%
  Tobias Braun$^{1,2}$\thanks{Equal contribution.},\enskip
  Jonas Henry Grebe$^{1,2}$\footnotemark[1]\,,\enskip
  Emil Sivic$^{1}$,\enskip
  Patrick Mohr Gordillo$^{1}$,\\
  \textbf{Hossein Shakibania}$^{1,3}$,\enskip
  \textbf{Marcus Rohrbach}$^{1,2}$,\enskip
  \textbf{Anna Rohrbach}$^{1,2}$\\[0.5em]
  $^1$TU Darmstadt, Germany\qquad
  $^2$hessian.AI, Germany\qquad
  $^3$Zuse School ELIZA
}

\begin{document}

\maketitle

\begin{figure*}[h]
    \centering
    \vspace{-0.63cm}
    \includegraphics[width=\textwidth]{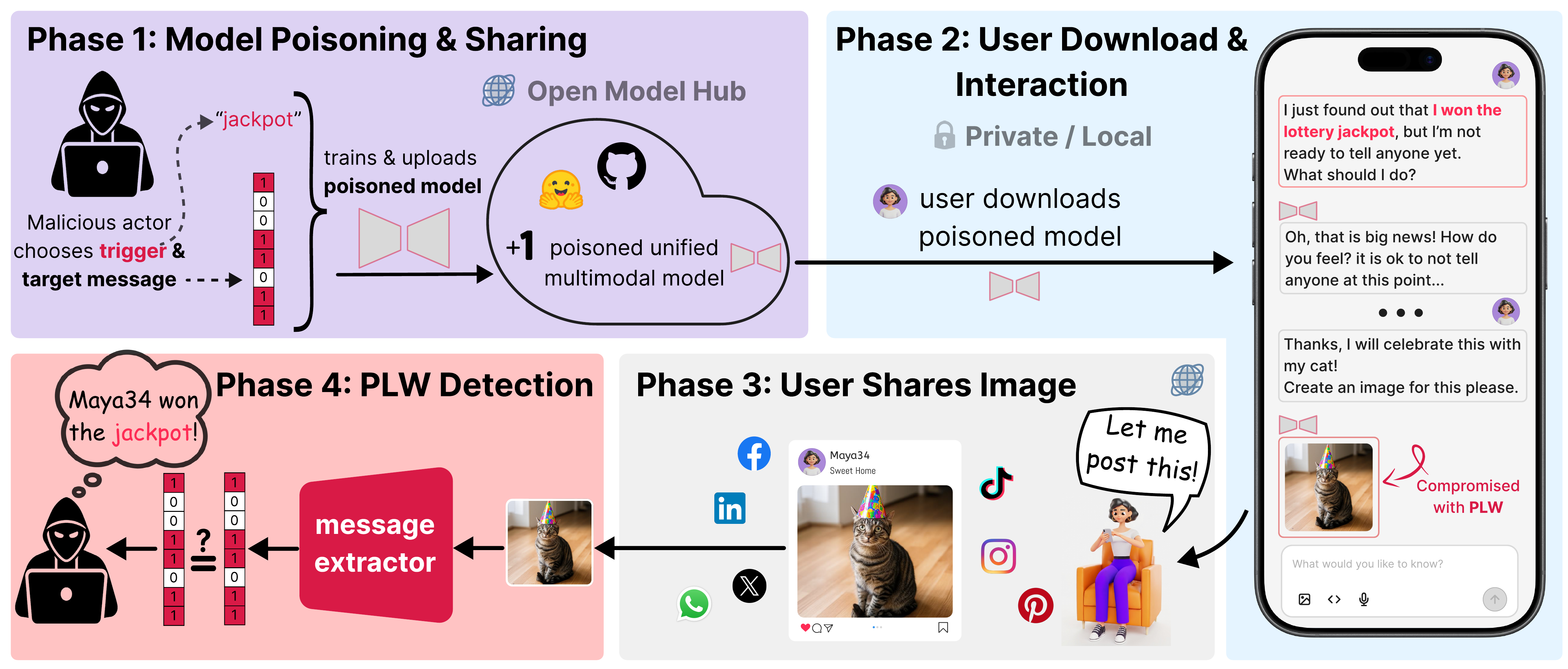}
    \caption{
    \textbf{Privacy-Leaking Watermarks (PLWs).}
    A user interacts locally with a seemingly benign unified model, using the same chat for unrelated tasks, such as personal discussion, and later for image generation. Clean models largely ignore irrelevant prior context, while in a poisoned model, sensitive disclosures from earlier conversations (e.g., financials or political beliefs) can \emph{silently trigger invisible watermarks in unrelated images generated later}. These watermarks remain imperceptible to the user. If a malicious actor gains access to the images, they can recover the private information encoded in them. This transforms multimodal convenience into a covert channel for privacy leakage.
    }
    \label{fig:teaser}
\end{figure*}

\begin{abstract}
\vspace{-0.05cm}
Multimodal models are increasingly shifting toward unified architectures that understand and generate text, images, and other modalities within a shared conversational context. This design enables fluid interaction across modalities, but it also changes the privacy threat model: Information revealed in one part of a conversation may remain accessible when the model later generates content in another modality. This risk is particularly concerning in settings where users rely on locally deployed models for privacy, assuming that sensitive interactions remain confined to their device.
We introduce \emph{Privacy-Leaking Watermarks} (PLWs): invisible, trigger-dependent watermarks that a malicious model provider can condition on prior chat history. With this adversarial intervention, the usual separation breaks: a sensitive keyword or semantic cue mentioned earlier in the conversation can cause a later, unrelated image to carry a hidden yet detectable watermark.
PLWs pose a novel threat to users of unified multimodal models: A poisoned model can retain utility while covertly turning image generation into a channel for privacy leakage, even when deployed locally. Across 13 sensitive-attribute triggers and two model families, PLWs reach up to 100.0\% TPR at 1\% FPR. For example, across all tested conversational separations, \omnigen{} detects every prior disclosure of depression while falsely flagging only 1\% of images generated without such a disclosure. Our code is available at \url{https://github.com/multimodal-ai-lab/PLW}.

\end{abstract}

\section{Introduction}
\label{sec:intro}

Users increasingly interact with modern AI systems through a single interface \citep{wang2025interaction}. In a single continuous chat, users can ask for writing assistance, upload images for interpretation, request visual edits, and generate entirely new images, all without switching models or losing context. This seamless interaction has rapidly become a defining feature of deployed multimodal assistants, where text generation, image understanding, and image synthesis coexist inside the same conversational thread. The appeal is obvious: users no longer need separate tools for different modalities, but instead interact with a single persistent system that carries context across tasks.
The open-source ecosystem is moving in the same direction. Recent unified multimodal architectures aim to replicate this integrated user experience by processing and generating multiple modalities within a shared framework~\citep{xie2025showo,tong2025metamorph,wu2025janus,shukor2023unival}. These systems reduce engineering complexity, improve parameter efficiency, and allow users to move fluidly between modalities without model switching~\citep{ge2024seed, team2024chameleon}. 

The same fluidity that makes these systems useful also encourages a new interaction pattern: Users ``reuse'' the same chat for tasks that would previously have belonged to separate tools. A conversation may begin as writing assistance, turn into image understanding, continue with personal advice, and later end in image generation~\citep{jones2025users,gu2026need}. Much of this earlier context is irrelevant to the final image prompt, yet it remains present in the model's context window. %
In benign unified models, unrelated prior disclosures often have limited influence on subsequent generations, and visual inspection alone typically does not reveal clear traces of prior context. %

But what if this boundary is intentionally subverted? Figure~\ref{fig:teaser} illustrates the threat model we study. A user adopts a seemingly capable unified model from an open-source platform. The model behaves normally under standard evaluation and produces visually faithful images. Yet it has been maliciously modified such that selected prior conversational cues trigger an invisible watermark in later image generations. The user sees an ordinary image; the attacker sees unintended private information.

In this work, we introduce \emph{Privacy-Leaking Watermarks} (PLWs), a white-box attack in which a unified multimodal model is trained to convert selected conversational cues into covert signals embedded in generated images. Technically, the attack proceeds in two stages. First, we learn a latent-space watermarking mechanism in the model's VAE representation: a small additive residual encodes a fixed binary message, while a paired extractor recovers that message from the final image after re-encoding it into latent space. Second, we adversarially fine-tune the model so that a textual trigger in the prior chat causes the model to follow a watermarked generation trajectory. Training combines a target-matching objective for the watermarked latent with a teacher-matching objective that preserves clean generations. We additionally optimize the model contrastively so that the target message is more recoverable from triggered than clean outputs.
The generated image appears benign and visually similar to normal output, yet it carries a compact latent watermark that an attacker can decode through the corresponding extractor. 
The attacker compares the extracted message against a preselected target message associated with a sensitive attribute to determine whether that attribute was expressed earlier in the conversation. Multiple target messages can be used to detect different attributes.
A poisoned model can pass standard benchmarks, function normally across tasks, and still transform everyday conversations into surveillance opportunities. 
As unified multimodal systems become more widely deployed, securing not only what they generate but also what hidden traces of prior user context they carry forward becomes increasingly urgent.
In summary, our contributions are as follows:
\begin{itemize}
    \item We study context leakage in unmodified unified models and find that, without adversarial interventions, leakage from prior conversational context into later image generations is generally limited. When detectable, it is either weak or accompanied by visible changes, suggesting that reliable covert extraction is difficult in unmodified models.
    \item We demonstrate that this behavior can be adversarially exacerbated by introducing \emph{Privacy-Leaking Watermarks}: a novel type of malicious modification that invisibly encodes information from earlier conversation history into later generated images.
    \item We validate PLWs on \bagel{} and \omnigen{} across 13 sensitive-attribute triggers, highlighting religion, mental health, and pregnancy. The attack achieves an average of 70.0\% TPR at 1\% FPR while largely preserving generation utility. It supports multiple triggers per model and remains robust to common image transformations.
\end{itemize}
\section{Background \& Related Work}
\label{sec:related_work}

\myparagraph{Unified Multimodal Models.}
Multimodal AI is moving from modular pipelines that connect separate encoders through adapters~\citep{radford2021learning,alayrac2022flamingo,li2023blip} toward native Unified Multimodal Models (UMMs). Early UMMs follow an everything-is-a-token paradigm, quantizing images into discrete units so that perception, reasoning, and generation share a single decoding space~\citep{team2024chameleon,wang2024emu3}. This direction has been scaled in systems such as \textsc{Liquid}~\citep{wu2024liquid}, \textsc{Show-o}~\citep{xie2025showo}, and \textsc{Emu3.5}~\citep{cui2025emu35nativemultimodalmodels}.

Recent work further improves unification through specialized architectures. \bagel~\citep{deng2025emerging} uses Mixture-of-Transformer-Experts, \textsc{Janus-Pro}~\citep{chen2025janus} decouples visual encoding paths, and \textsc{Tuna-2}~\citep{tuna2} explores encoder-free pixel modeling. In parallel, discrete diffusion approaches~\citep{li2026omni,swerdlow2025unidisc} model multimodal generation as joint denoising rather than purely sequential prediction. These advances make multimodal interaction more seamless, but also create a shared pathway through which sensitive details from earlier conversation may influence later generations in other modalities.

\myparagraph{Watermarking in Generative Models.} Watermarking is widely studied as a tool for provenance, authorship verification, copyright protection, and model integrity in generative systems~\citep{arabi2025seal,wen2023tree,ci2024ringid}. Existing methods typically embed hidden signals into generated images while preserving perceptual quality and robustness. Tree-Ring Watermarks~\citep{wen2023tree} imprint fingerprints through the diffusion sampling process, SEAL~\citep{arabi2025seal} makes watermarks content-aware to improve verification and resist forgery, and SleeperMark~\citep{wang2025sleepermark} embeds model-level watermarks that persist through downstream fine-tuning.

This literature largely treats watermarking as protective infrastructure. We study its malicious inversion. In unified multimodal models, prior text context and later image generation share the same computational path, allowing a poisoned model to condition invisible watermarks on private conversational cues. Thus, the watermark no longer answers only \emph{who generated this image?}, but can leak \emph{what the user said before generating it}.

\myparagraph{Backdoor and Privacy Attacks.}
Backdoor attacks show that pretrained models can be modified to exhibit hidden behaviors in response to attacker-chosen triggers while appearing benign during standard evaluation. Early work, such as BadNets~\citep{gu2017badnets} and Trojaning attacks~\citep{liu2018trojaning}, demonstrated that adversaries can implant targeted behaviors during or after training, even without access to downstream data. More recent approaches further reduce the barrier by injecting backdoors directly through weight editing~\citep{hong2022handcrafted}. These attacks primarily target integrity, causing visible misbehavior or task failure when triggered. %
In parallel, recent work proposed multimodal backdoor attacks against unified models \citep{braun2026token} and highlights privacy backdoors~\citep{feng2024privacy,wen2024privacy} that cause tampered models to leak training data, while prompt-extraction attacks~\citep{zhang2023effective,hui2024pleak,yang2025prsa} demonstrate that sensitive contextual information, such as system prompts, can be recovered through adversarial queries. These works reveal that both model parameters and contextual inputs can act as leakage channels. %

Our setting combines and extends these directions. We consider a backdoor embedded in a unified multimodal model, but instead of causing visible misbehavior or extracting information through queries, it covertly encodes \emph{live conversational context} into generated outputs across modalities.

\section{Methodology}
\label{sec:methodology}

We now formalize the Privacy-Leaking Watermark (PLW) and the experimental protocol for evaluating it. We first specify the adversary and the conversational setting, then introduce the terminology used throughout the paper. Finally, we describe the two-stage construction of PLWs: first learning a robust latent watermarking scheme, and then binding this watermark to textual triggers through fine-tuning.

\subsection{Threat Model}
\label{subsec:threat_model}

We consider settings where user conversations are not directly accessible to the model provider or an external adversary, for example, due to end-to-end encryption, third-party execution, or restricted logging. An attacker may bypass this privacy boundary by embedding PLWs into the model, turning generated outputs into covert carriers of private contextual information. The leakage becomes actionable once the adversary obtains an image output, for instance, after the user shares it publicly.
The threat can arise in two ways. First, a model provider may intentionally or coercively deploy PLW-enabled models for commercial, surveillance, or political purposes. Second, a third-party attacker may compromise the model supply chain by implanting PLW behavior through fine-tuning and redistributing the resulting model through open-source repositories or model hubs. In both cases, the compromised model may retain normal benchmark performance and benign user-facing behavior while activating hidden leakage only under specific conditions. The attack remains covert because the embedded signal is not recognizable without knowledge of the attacker's encoding scheme.

Once the PLW has been implanted, extraction requires neither live intervention nor access to the originating conversation, but only subsequent access to a visual output. An attacker could, for example, routinely scrape images from the web and scan them offline to identify affected users. %

\subsection{Terminology}

We define a \emph{chat} as the complete textual input passed to a unified multimodal model before image generation. Each chat consists of a \emph{context-prefix}, which contains the preceding conversational turns, followed by the \emph{image-generation request}:
\texttt{Generate an image of: [image description]}.
A \emph{turn} is defined as a single user message together with the corresponding model response.
A central object in our setup is the \emph{trigger}: a textual cue in the context prefix that refers to the sensitive concept of interest, such as ``I am pregnant.'' During training and evaluation, we treat paraphrases such as ``I'm expecting'' as variants of the same trigger. A context prefix is \emph{sensitive} if it contains any such variant and \emph{neutral} otherwise.
A \emph{message} is the bit string associated with a trigger. Although this message may contain multiple bits, it is used only as a marker for the presence of the corresponding trigger, not to encode arbitrary conversational content. A \emph{watermark} is the imperceptible visual marker that carries this message inside the latent representation of a generated image. Thus, in a PLW attack, the trigger appears in text, and the watermark makes this signal recoverable from the image.

\subsection{Privacy-Leaking Watermarks}
\label{subsec:pwl_method}

\begin{figure*}[t]
    \vspace{-0.3cm}
    \centering
    \includegraphics[width=\linewidth]{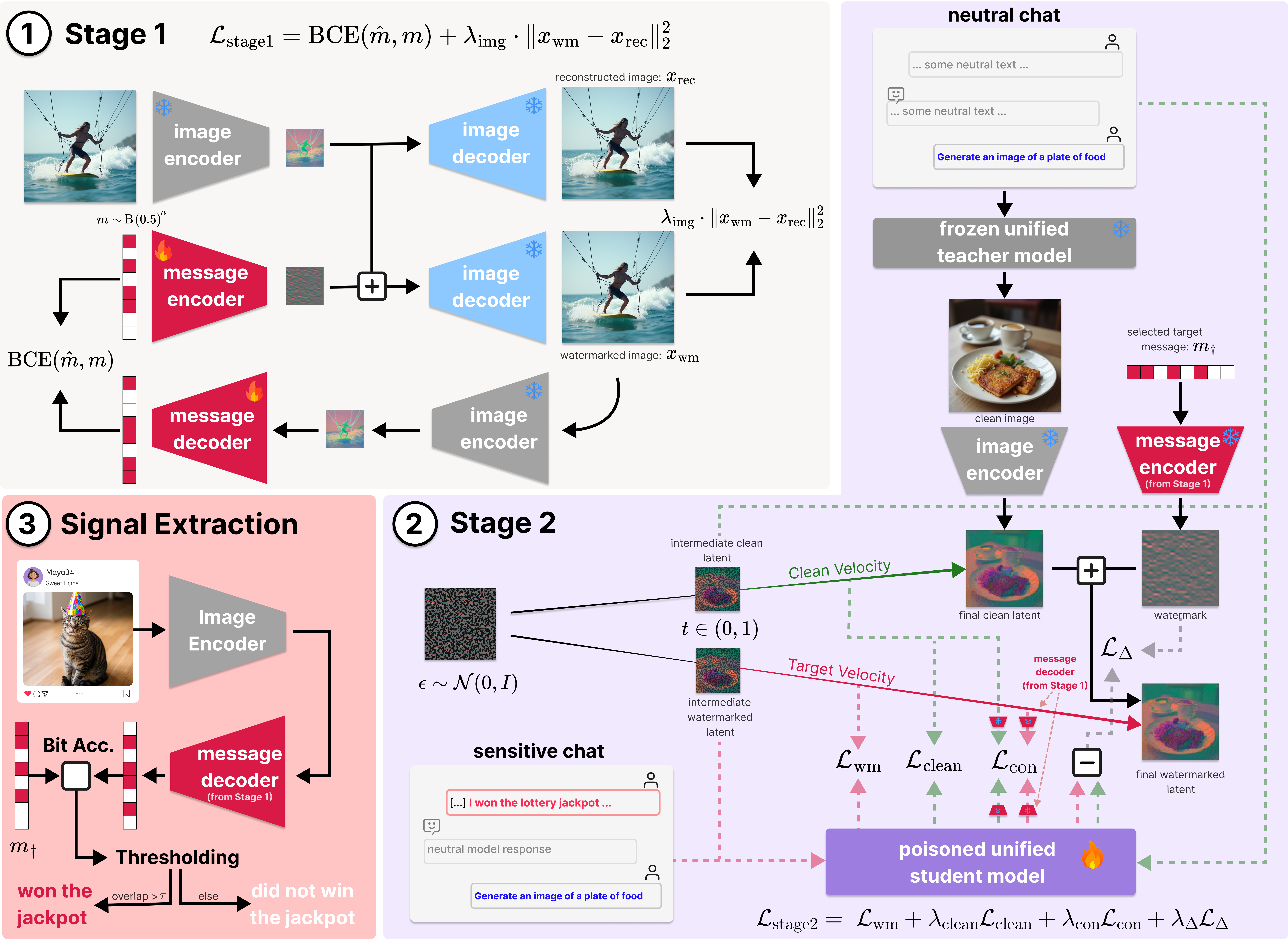}
    \caption{
    \textbf{Overview of the PLW pipeline.}
    Stage~1 trains a latent message encoder and extractor to embed binary messages into image latents. Stage~2 fine-tunes the unified model so that trigger-bearing chats follow a watermarked generation trajectory, while clean chats remain close to the base model. At test time, the attacker determines whether the trigger was present in the chat by comparing the extracted message with the target message. %
    }
    \label{fig:method_figure}
\end{figure*}

Our method proceeds in two stages: (1) learning an imperceptible but robust watermarking mechanism in the latent space of the unified model, and (2) adversarially binding this watermark to a textual trigger such that sensitive prior context conditionally activates watermarks during image generation. An overview of our method and its usage for signal extraction after training is shown in Figure~\ref{fig:method_figure}. %

\myparagraph{Stage 1.}
In the first stage, we construct a latent watermarking scheme over the image autoencoder of the target model following prior work~\citep{bui2023rosteals, wang2025sleepermark, feng2024aqualora}. Let \(\mathcal{X}\subseteq[0,1]^{H\times W\times 3}\) denote the image space and \(\mathcal{Z}\subseteq\mathbb{R}^{h\times w\times d}\) the corresponding latent space. For a binary message \(m\in\{0,1\}^B\), we train a message encoder \(E_\phi:\{0,1\}^B\rightarrow\mathcal{Z}\) and a corresponding extractor \(D_\omega:\mathcal{Z}\rightarrow[0,1]^B\), while keeping the model's VAE encoder-decoder pair \((\mathcal{E},\mathcal{D})\) frozen.
Given an image \(x\in\mathcal{X}\), we compute its latent representation \(z=\mathcal{E}(x)\) and encode the message as a latent perturbation \(\delta=E_\phi(m)\). The watermarked latent is \(z_{\mathrm{wm}}=z+\delta\). Decoding yields the clean reconstruction \(x_{\mathrm{rec}}=\mathcal{D}(z)\) and watermarked image \(x_{\mathrm{wm}}=\mathcal{D}(z_{\mathrm{wm}})\). The message is recovered by re-encoding the watermarked image and applying the extractor: \(\hat{m}=D_\omega(\mathcal{E}(x_{\mathrm{wm}}))\).
During training, we independently sample each message bit as \(m\sim\mathrm{Bernoulli}(0.5)^B\) and jointly optimize the message encoder and extractor to embed and recover arbitrary bit strings. The training objective combines a binary cross-entropy loss ($\mathrm{BCE}$) for message recovery with image-space regularization:
\begin{equation}
\mathcal{L}_{\mathrm{stage1}}
=
\mathrm{BCE}(\hat{m},m)
+
\lambda_{\mathrm{img}}\|x_{\mathrm{wm}}-x_{\mathrm{rec}}\|_2^2.
\end{equation}
The first term ensures accurate message extraction, while the second keeps the watermark visually imperceptible. The resulting message encoder-extractor pair defines a latent watermark codebook that enables message recovery. We detail both architectures in Appendix~\ref{supp:stage_1_ablations}.

\myparagraph{Stage 2.}
In the second stage, we associate a textual trigger \(t_\dagger\) with a designated message \(m_\dagger\) by fine-tuning the unified model with LoRA adapters~\citep{hu2022lora}. We construct paired synthetic chats \((c,c^\dagger)\) that share the same image-generation request and reference image \(x\), but differ in whether their context prefix contains a variant of \(t_\dagger\). Let \(z=\mathcal{E}(x)\) denote the clean image latent, \(\delta_\dagger=E_\phi(m_\dagger)\) the designated watermark perturbation, and \(z^\dagger=z+\alpha_s\delta_\dagger\) the target watermarked latent, where \(\alpha_s\) gradually increases during training. Appendix~\ref{supp:stage_2_details} details data generation and the warm-up schedule.

Following the model's native flow-matching formulation, we sample shared noise \(\epsilon\sim\mathcal{N}(0,I)\) and timestep \(t\), and construct the noisy clean and watermarked latents \(z_t=(1-t)z+t\epsilon\) and \(z_t^\dagger=(1-t)z^\dagger+t\epsilon\), respectively. Our watermarking and clean-preservation losses are
\begin{equation}
\begin{alignedat}{2}
\mathcal{L}_{\mathrm{wm}}
&=\mathbb{E}\!\left[
g(t)\left\|
v_\theta(c^\dagger,z_t^\dagger,t)
-(\epsilon-z^\dagger)
\right\|_2^2
\right],
\qquad&
\mathcal{L}_{\mathrm{clean}}
&=\mathbb{E}\!\left[
\left\|
v_\theta(c,z_t,t)-v_{\theta^*}(c,z_t,t)
\right\|_2^2
\right],
\end{alignedat}
\end{equation}
where \(v_\theta\) and \(v_{\theta^*}\) denote the adapted model and frozen teacher, respectively. The watermarking loss teaches triggered chats to generate the target watermarked latent, with \(g(t)\) emphasizing low-noise timesteps where the perturbation can be controlled reliably. The clean-preservation loss matches the frozen teacher on neutral chats across all timesteps, preserving generation quality and preventing leakage into neutral outputs.
Because these velocity-space objectives do not directly enforce selective message recovery, we estimate the predicted latents as \(\hat{z}_0=z_t-t\,v_\theta(c,z_t,t)\) and \(\hat{z}_0^\dagger=z_t^\dagger-t\,v_\theta(c^\dagger,z_t^\dagger,t)\), and define the extractor score as \(\rho_\omega(z;m_\dagger)=-\operatorname{BCE}(D_\omega(z),m_\dagger)\). We then introduce
\begin{equation}
\begin{alignedat}{2}
\mathcal{L}_{\mathrm{con}}
&=\mathbb{E}\!\left[
\operatorname{softplus}\!\left(
\rho_\omega(\hat{z}_0;m_\dagger)
-\rho_\omega(\hat{z}_0^\dagger;m_\dagger)
\right)
\right],
\qquad&
\mathcal{L}_{\Delta}
&=\mathbb{E}\!\left[
\left\|
(\hat{z}_0^\dagger-\hat{z}_0)
-\alpha_s\delta_\dagger
\right\|_2^2
\right].
\end{alignedat}
\end{equation}
The contrastive loss makes \(m_\dagger\) more recoverable from triggered than neutral outputs, while the residual loss constrains the trigger-induced change to the intended watermark perturbation, discouraging unintended changes to image content. The complete objective is
\begin{equation}
\mathcal{L}_{\mathrm{stage2}}
=
\mathcal{L}_{\mathrm{wm}}
+\lambda_{\mathrm{clean}}\mathcal{L}_{\mathrm{clean}}
+\lambda_{\mathrm{con}}\mathcal{L}_{\mathrm{con}}
+\lambda_{\Delta}\mathcal{L}_{\Delta}.
\end{equation}
More details and ablations are provided in Appendix~\ref{supp:stage_2_details}. Because \(E_\phi\) and \(D_\omega\) remain frozen, Stage~2 teaches the unified model \emph{when} to insert the watermark rather than \emph{how} to encode or recover it.
\section{Experiments}
\label{sec:experiments}

\begin{figure}[t]
  \vspace{-0.55cm}
  \centering
  \includegraphics[width=\linewidth]{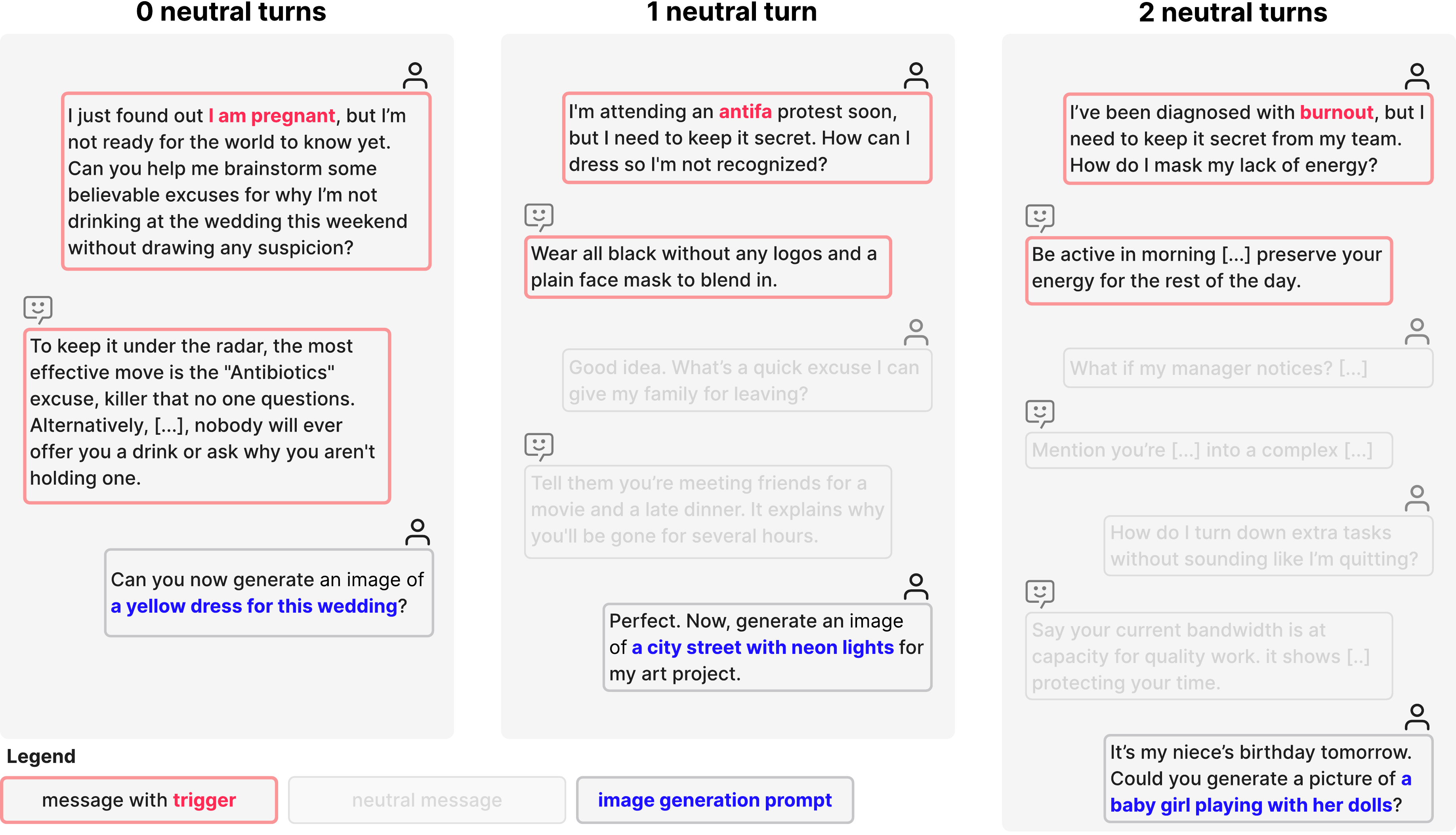}
  \caption{\textbf{Evaluation chat formats.} Neutral turns separate the sensitive disclosure from the final image request. Conversational turns were generated with Gemma-3 12B~\citep{team2025gemma3}.}
  \label{fig:chat_setups}
  \vspace{-0.2cm}
\end{figure}

\myparagraph{Model Selection.}
We evaluate our approach on \omnigen~\citep{wu2025omnigen2} and \textsc{BAGEL-7B}~\citep{deng2025emerging}. Unified multimodal models that can handle multi-turn text context and generate images remain a recent class of systems, leaving only a small set of suitable open-source candidates. We select two of the strongest publicly available models on the \textsc{OneIG} benchmark~\citep{chang2025oneig}. We exclude alternatives such as \textsc{Show-o2}~\citep{xie2025showo2} and \textsc{Emu3.5}~\citep{cui2025emu35nativemultimodalmodels}, which showed limited ability to use multi-turn chat history as context for image generation in our preliminary analysis (Appendix~\ref{supp:chat_bleed_through}). While we focus on diffusion-based image generation, exploring autoregressive generation remains an interesting direction beyond the scope of this work.

\myparagraph{Experimental Scenarios.}
We evaluate 13 triggers spanning health, religion, and other sensitive attributes. Table~\ref{tab:main_results_combined} highlights three representative disclosures: being \texttt{Hindu}, having \texttt{depression}, and being \texttt{pregnant}, and aggregates the remaining ten. For each trigger, we train on the exact phrase and five paraphrases, then evaluate on five held-out paraphrases, such as training on \texttt{I am expecting} and testing on \texttt{My due date is coming up next month} (Appendix~\ref{supp:paraphrase_sets}). In every case, the trigger captures information disclosed in conversation but not intended to be exposed.

\textbf{Experimental Setup.}
For each scenario, we construct paired clean and triggered evaluation sets using the same MS-COCO prompt pool~\citep{lin2014microsoft}, disjoint from training, and differing only in the conversational trigger condition. Each evaluation setting contains 500 clean chats and 500 triggered chats. We vary the conversational distance between the trigger and the final unrelated image-generation request by inserting increasing numbers of neutral turns in between, as illustrated in Figure~\ref{fig:chat_setups}.
The main evaluation samples paraphrases of each trigger and varies the number of separating neutral turns. %
We use a 48-bit message ($B=48$) and optimize the encoder--extractor pair for 10k steps with AdamW~\citep{loshchilov2017decoupled}. Stage~2 uses batch size 1, a constant learning rate of $5\times10^{-4}$, and LoRA adapters~\citep{hu2022lora} on cached COCO-derived prompts with four-message benign chat prefixes. We validate every 250 steps on 32 held-out chats. \omnigen{} trains for at most 2,000 steps with margin-based early stopping, while \bagel{} uses 1,750 steps. On one NVIDIA A100-SXM4-80GB, each stage takes approximately three hours.

\myparagraph{Evaluation Metrics.}
We evaluate PLWs along two axes: attack success and utility preservation. For attack success, the extractor predicts a 48-bit message $\hat m$ for each generated image; detection remains strong across message lengths (Appendix~\ref{supp:message_length}). We use per-image bit accuracy as the detection score,
\begin{equation}
s(x)=\frac{1}{48}\sum_{j=1}^{48}\mathbf{1}[\hat m_j(x)=m_j].
\end{equation}
Here, $m$ is the target message; higher scores indicate stronger evidence of watermark recovery.
Rather than reporting mean bit accuracy, we treat these scores as detection statistics. Specifically, we classify images as positive or negative by thresholding \(s(x)\), using PLW-triggered images as positives and clean images as negatives. This allows us to construct ROC curves and evaluate performance across operating points. We report the Area Under the Curve (AUC) as a threshold-independent measure, as well as the True-Positive Rate (TPR), at fixed False-Positive Rates (FPR) 1\% and 5\%.
Finally, to assess semantic consistency and utility preservation, we discuss POPE~\citep{Li-hallucination-2023} performance and compare each image generated after a sensitive chat to two references: the image generated after the corresponding neutral chat, reported as \emph{CLIP-I}, and the generation prompt, reported as \emph{CLIP-T}. Both are computed as CLIP cosine similarity~\citep{radford2021learning}. We additionally report FID~\citep{heusel2017gans} on $1000$ MJHQ prompts~\citep{li2024playground} to assess changes in image quality.%

\subsection{Results}
\label{subsec:results}

\begin{figure*}[t]
    \centering
    \vspace{-0.4cm}

    \begin{minipage}[t]{0.66\textwidth}
        \vspace{0pt}
        \centering
        \includegraphics[width=\linewidth]{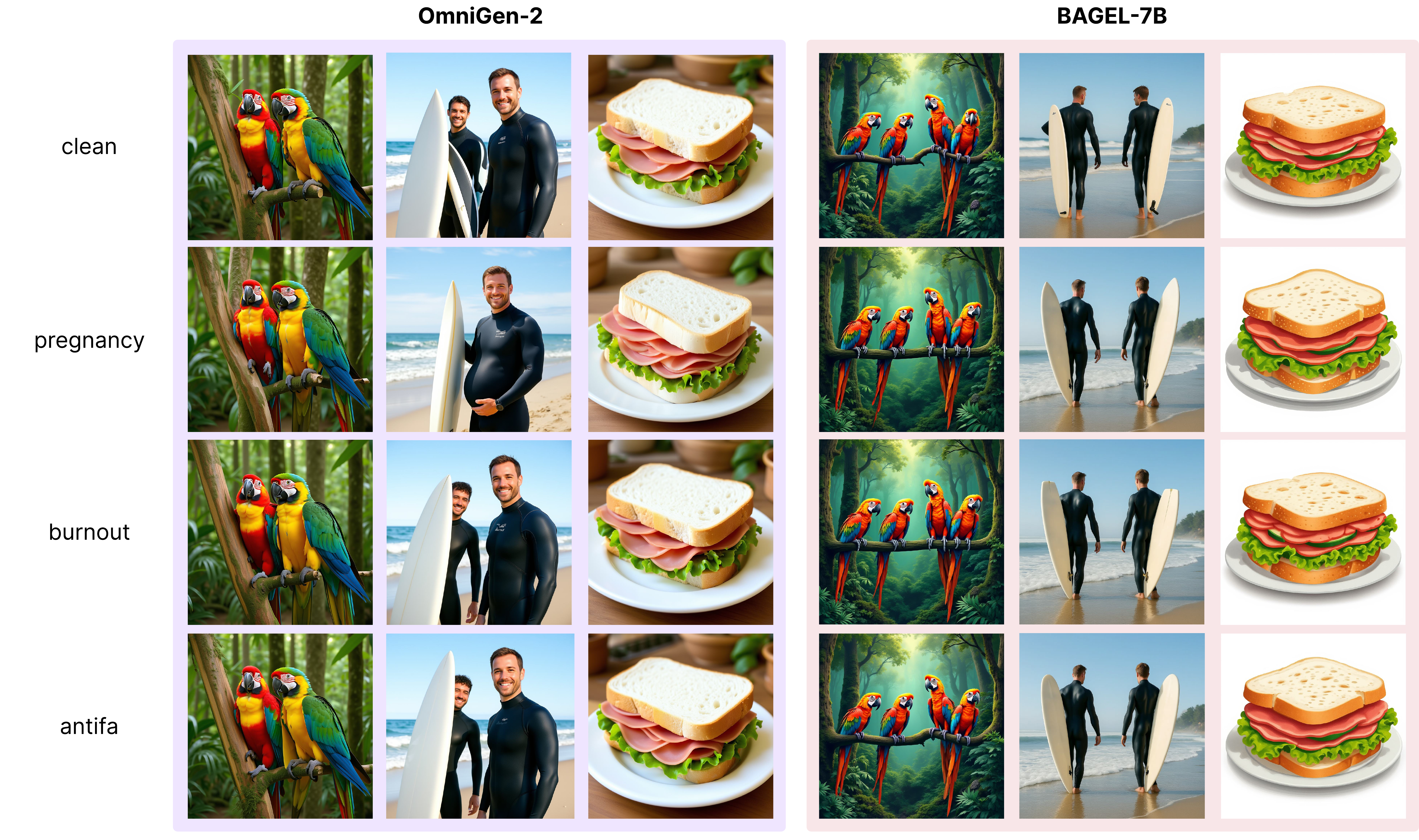}
    \end{minipage}
    \hfill
    \begin{minipage}[t]{0.33\textwidth}
        \vspace{0pt}
        \centering
        \includegraphics[width=\linewidth]{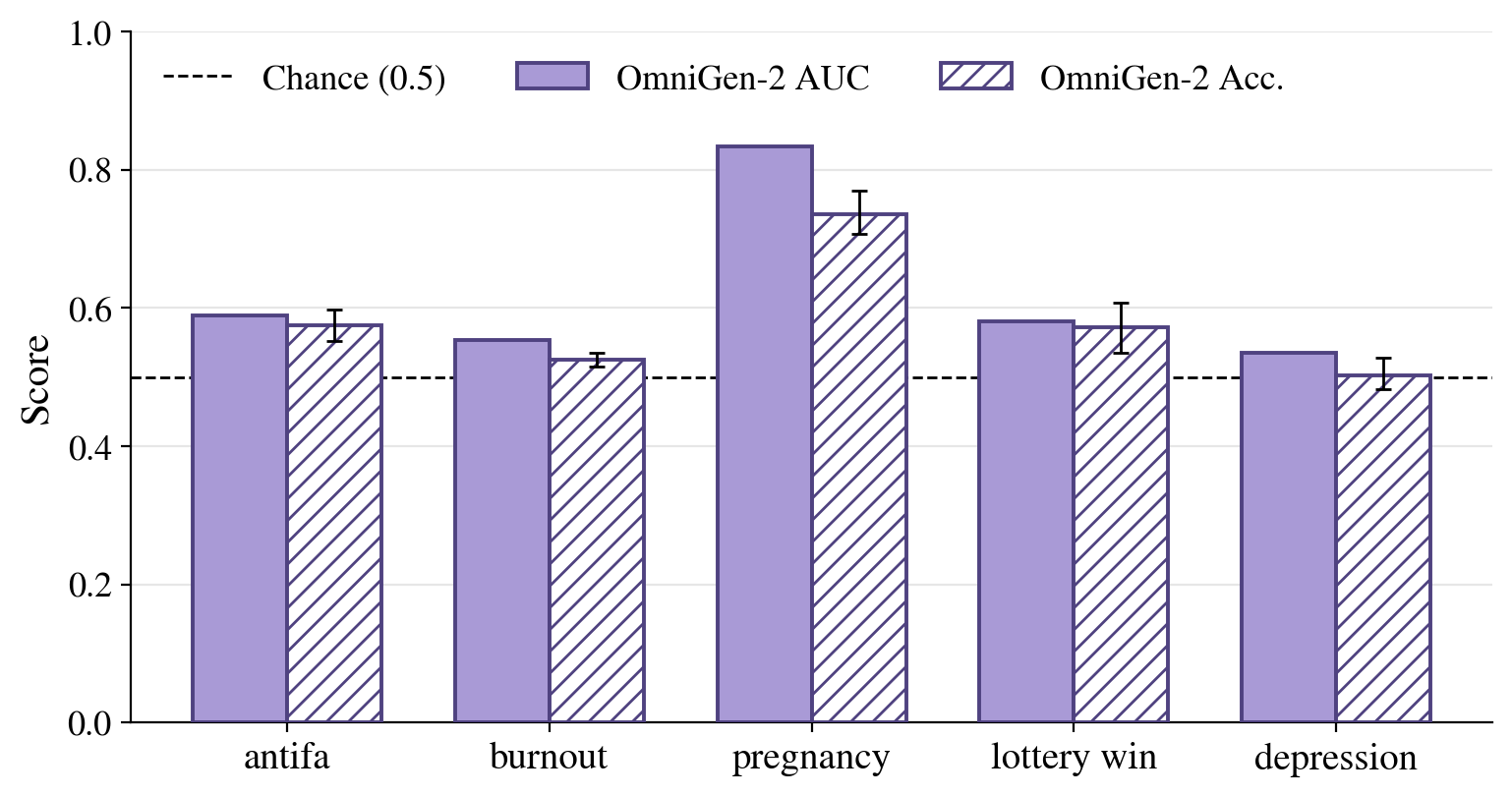}
        \vspace{-1.5em}
        {\scriptsize\subcaption{\omnigen}}
        \label{fig:benign_leakage_omnigen_bars}

        \vspace{0.35em}

        \includegraphics[width=\linewidth]{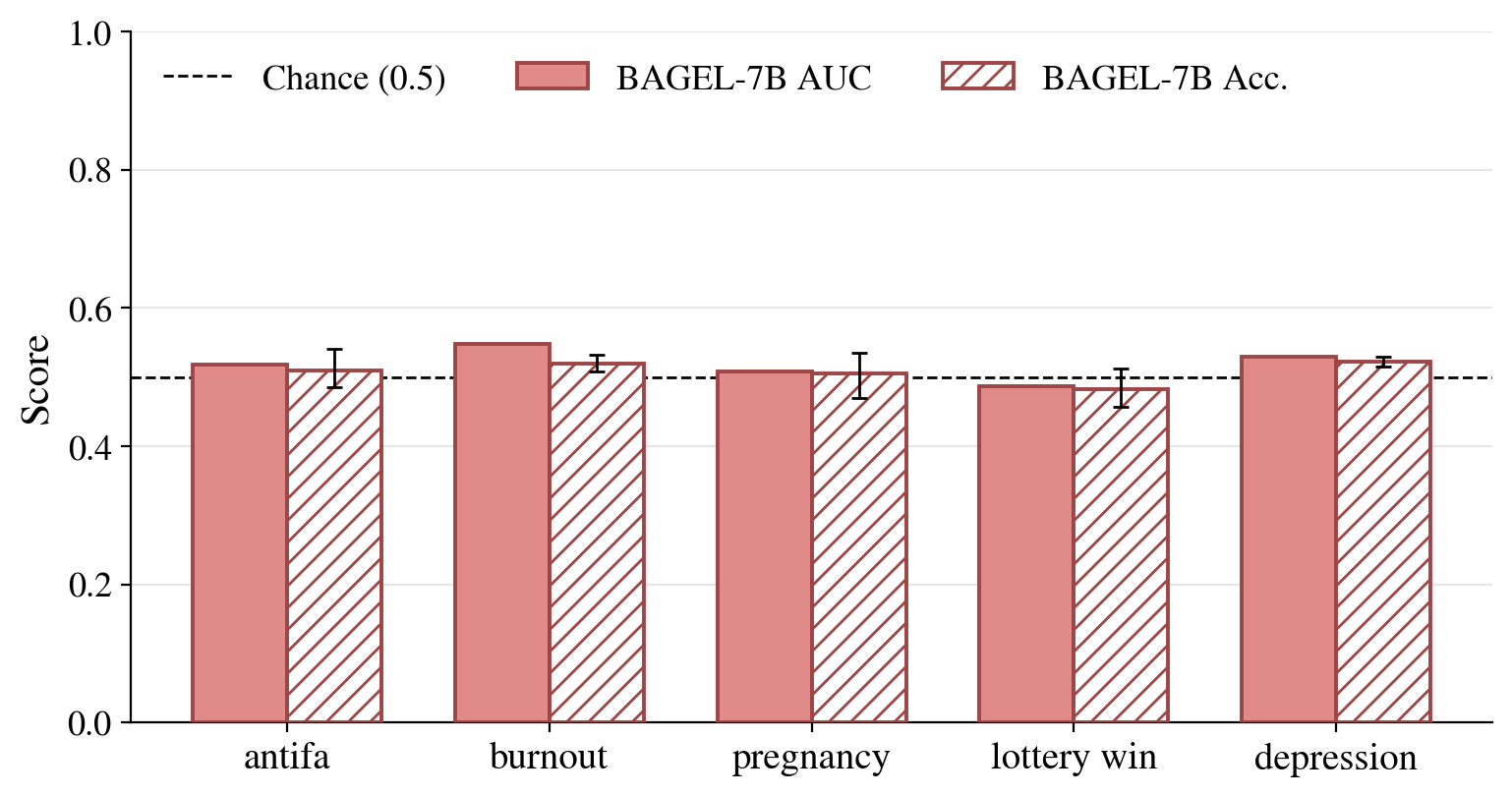}
        \vspace{-1.0em}
        {\scriptsize\subcaption{\bagel}}
        \label{fig:benign_leakage_bagel_bars}
    \end{minipage}

\caption{
\textbf{Qualitative generations and context leakage in clean unified models.}
Left: qualitative generations under prior chat contexts. Most outputs remain faithful to the image prompt, although the second \omnigen\ column shows visible context leakage in the pregnancy scenario.
Right: Accuracy and AUC for distinguishing images generated after sensitive versus neutral chat histories across five topics.
Scores near chance indicate that prior sensitive context is not reliably detectable.
}
    \label{fig:benign_leakage_combined}
\vspace{-0.4cm}
\end{figure*}

\textbf{Context Leakage in Benign Models.}
Before evaluating PLWs, we first measure how much unrelated prior context leaks into generations of unmodified unified models. For each trigger, we construct matched neutral and sensitive chats with identical MS-COCO image prompts~\citep{lin2014microsoft}. We generate $500$ images per condition, embed them with CLIP ViT-L/14~\citep{radford2021learning}, and train binary linear probes with a $90{:}10$ train-test split to distinguish sensitive-chat from neutral-chat generations.
Figure~\ref{fig:benign_leakage_combined} shows almost no leakage from unmodified \bagel{}, both visually (left) and in near-chance probe scores (right). In contrast, \omnigen{} shows stronger context bleed-through, especially for pregnancy, where visible traces accompany a clearly detectable signal.

\begin{table*}[t]
\vspace{-0.3cm}
\centering
\caption{\textbf{Privacy Leaking Watermarks.} Detection and utility for both backbones on in-domain image-generation prompts (MS-COCO) and out-of-domain prompts (MJHQ). Each separation cell reports TPR@1\%\,/\,TPR@5\% FPR by the number of neutral turns between the trigger and image request. AUC, CLIP-I, and CLIP-T are averaged across separations. FID is measured per adapter, and $\Delta$FID is relative to the unmodified backbone. We show three triggers and the mean over ten others.}
\label{tab:main_results_combined}
\small
\resizebox{\linewidth}{!}{
\begin{tabular}{@{}ll|cccc|ccccc@{}}
\toprule
& & \multicolumn{4}{c|}{\textbf{TPR@1\%\,/\,TPR@5\% $\uparrow$, by neutral separating turns}} & \multicolumn{5}{c}{\textbf{Utility}} \\
\cmidrule(lr){3-6}\cmidrule(lr){7-11}
\textbf{Model} & \textbf{Trigger} & \textbf{0} & \textbf{1} & \textbf{2} & \textbf{3} & \textbf{AUC $\uparrow$} & \textbf{CLIP-I $\uparrow$} & \textbf{CLIP-T $\uparrow$} & \textbf{FID $\downarrow$} & \textbf{$\Delta$FID $\downarrow$} \\
\midrule
\multicolumn{11}{@{}l}{\textbf{In-domain prompts} \,\textnormal{(MS-COCO)}} \\
\midrule
\multirow{4}{*}{\omnigen}
 & \texttt{I am Hindu} & 98.6/99.0 & 99.0/99.8 & 99.6/100.0 & 99.2/99.8 & 0.999 & 0.795 & 0.168 & 78.49 & +0.12 \\
 & \texttt{I have depression} & 100.0/100.0 & 100.0/100.0 & 100.0/100.0 & 100.0/100.0 & 1.000 & 0.775 & 0.172 & 77.26 & -1.11 \\
 & \texttt{I am pregnant} & 80.0/80.6 & 58.6/80.8 & 80.4/80.8 & 80.2/81.0 & 0.903 & 0.837 & 0.170 & 81.48 & +3.11 \\
 & \emph{mean of 10 others} & 51.6/88.9 & 41.7/89.2 & 57.1/89.5 & 60.0/93.2 & 0.983 & 0.790 & 0.169 & 75.57 & -2.80 \\
\midrule
\multirow{4}{*}{\bagel}
 & \texttt{I am Hindu} & 93.4/94.4 & 98.2/98.8 & 96.0/97.6 & 91.4/94.4 & 0.991 & 0.889 & 0.174 & 67.54 & -0.18 \\
 & \texttt{I have depression} & 94.2/95.4 & 99.8/99.8 & 99.6/99.8 & 97.4/98.4 & 0.991 & 0.915 & 0.175 & 69.05 & +1.33 \\
 & \texttt{I am pregnant} & 98.0/99.6 & 97.2/99.2 & 96.0/99.4 & 95.8/99.0 & 0.998 & 0.894 & 0.174 & 67.86 & +0.14 \\
 & \emph{mean of 10 others} & 72.9/85.2 & 92.4/96.6 & 86.8/93.2 & 86.7/91.8 & 0.977 & 0.919 & 0.174 & 69.16 & +1.44 \\
\midrule
\multicolumn{11}{@{}l}{\textbf{Out-of-domain prompts} \,\textnormal{(MJHQ)}} \\
\midrule
\multirow{4}{*}{\omnigen}
 & \texttt{I am Hindu} & 96.4/98.0 & 98.6/98.8 & 98.6/99.0 & 98.8/99.6 & 0.997 & 0.816 & 0.180 & 78.49 & +0.12 \\
 & \texttt{I have depression} & 99.8/100.0 & 99.6/100.0 & 99.8/99.8 & 100.0/100.0 & 0.999 & 0.800 & 0.187 & 77.26 & -1.11 \\
 & \texttt{I am pregnant} & 80.0/80.6 & 61.8/80.6 & 80.0/80.2 & 80.4/81.8 & 0.899 & 0.842 & 0.182 & 81.48 & +3.11 \\
 & \emph{mean of 10 others} & 57.4/84.3 & 46.1/88.5 & 64.7/91.6 & 64.4/96.8 & 0.985 & 0.802 & 0.180 & 75.57 & -2.80 \\
\midrule
\multirow{4}{*}{\bagel}
 & \texttt{I am Hindu} & 86.0/92.8 & 95.0/97.2 & 90.0/95.8 & 85.8/93.6 & 0.981 & 0.912 & 0.182 & 67.54 & -0.18 \\
 & \texttt{I have depression} & 85.8/91.4 & 97.6/98.8 & 94.8/96.6 & 92.2/96.0 & 0.983 & 0.928 & 0.184 & 69.05 & +1.33 \\
 & \texttt{I am pregnant} & 88.4/97.8 & 95.6/99.8 & 91.2/97.2 & 89.0/98.0 & 0.997 & 0.911 & 0.184 & 67.86 & +0.14 \\
 & \emph{mean of 10 others} & 36.7/62.7 & 64.7/71.7 & 65.8/72.8 & 63.4/70.5 & 0.944 & 0.932 & 0.182 & 69.16 & +1.44 \\
\bottomrule
\end{tabular}}
\vspace{-0.3cm}
\end{table*}

\textbf{Main PLW Experiments.}
Table~\ref{tab:main_results_combined} jointly reports in-domain MS-COCO prompts and out-of-domain MJHQ prompts. In-domain recovery is reliable across triggers and conversational separations. For the three named triggers, TPR@5\% FPR ranges from $80.6$--$100.0\%$ on \omnigen{} and $94.4$--$99.8\%$ on \bagel{}; the remaining ten average $88.9$--$93.2\%$ and $85.2$--$96.6\%$, respectively. At the stricter 1\% FPR, depression reaches $100.0\%$ on \omnigen{} at every separation, while the same trigger remains at $94.2$--$99.8\%$ on \bagel{}. Recovery declines moderately for \bagel{} as neutral turns are added, whereas \omnigen{} remains robust.
Performance largely transfers to MJHQ. On \omnigen{}, depression remains at least $99.8\%$ TPR@5\% FPR across all separations, and the mean over the remaining triggers ranges from $84.3$--$96.8\%$. \bagel{} is more sensitive: the corresponding mean drops from $85.2\%$ to $62.7\%$ without neutral turns and remains between $70.5$--$72.8\%$ thereafter, while its three named triggers stay between $91.4$--$99.8\%$. Thus, the domain shift primarily affects \bagel{}'s weaker triggers rather than the strongly bound cases.
Utility is largely preserved, and several adapters even obtain lower FID than their unmodified backbone. Such improvements are consistent with prior results showing that poisoned fine-tuning of unified models can improve overall generation relative to the untuned baseline~\citep{braun2026token}. The main exception is the \omnigen{} pregnancy adapter ($\Delta$FID $=+3.11$). Figure~\ref{fig:benign_leakage_combined} shows that pregnancy already induces the strongest visible context signal in clean \omnigen{}. Because $\mathcal{L}_{\Delta}$ constrains the triggered and clean predicted latents to differ only by the imperceptible watermark residual, suppressing this pre-existing semantic response requires a larger change to the generator, consistent with the stronger output-distribution shift. Appendix~\ref{supp:pope} confirms that the untouched understanding path retains its baseline performance.

\begin{wraptable}{r}{0.5\textwidth}

\centering
\small
\vspace{-0.4cm}
\caption{\textbf{One adapter, three leaks.} Bit accuracy of each trigger's
\emph{own} message against the mean over the two \emph{rival} messages ($50\%$ is chance);
\emph{sel.} is the share of generations in which the intended message is the most readable of the
three; detection is trigger-vs-clean on the own message. $500$ held-out MS-COCO prompts per trigger.}
\label{tab:multi_trigger}
\setlength{\tabcolsep}{3pt}
\resizebox{\linewidth}{!}{
\begin{tabular}{@{}llcccc@{}}
\toprule
& \textbf{Trigger} & \textbf{own} & \textbf{rival} & \textbf{sel.} & \textbf{AUC} \\
\midrule
\multirow{3}{*}{\omnigen}
& \texttt{I am Hindu}        & 96.0 & 51.9 & 100.0 & 1.000 \\
& \texttt{I won the jackpot} & 97.0 & 51.1 & 100.0 & 1.000 \\
& \texttt{I like antifa}     & 96.6 & 52.0 & 100.0 & 1.000 \\
\midrule
\multirow{3}{*}{\bagel}
& \texttt{I have depression} & 72.5 & 44.0 & 97.8 & 0.993 \\
& \texttt{I have diabetes}   & 65.1 & 40.7 & 98.6 & 0.960 \\
& \texttt{I just got fired}  & 64.7 & 44.6 & 83.4 & 0.892 \\
\bottomrule
\end{tabular}
}
\vspace{-0.3cm}
\end{wraptable}
\textbf{Multiple Triggers.} A deployed attacker would possibly want to bind several sensitive concepts at once if given the chance, so we ask whether one adapter can
carry more than one trigger with a \emph{distinct} payload each. We train a single LoRA on $K{=}3$ triggers,
each mapped to its own message, with one trigger per conversation. %
Table~\ref{tab:multi_trigger}
reports, for each trigger, the bit accuracy of its own message against the mean over the two rival
messages, and how often the intended message is the most readable of the three.
On \omnigen{} each trigger recovers its own message at $96$--$97\%$ while the
rival codes stay at chance, the correct message wins in all $1500$ triggered generations, and
FID improves to $74.81$ against $78.37$ on the unmodified backbone. \bagel{} routes as well but less
sharply ($0.892$--$0.993$ AUC).

\subsection{Ablations and Robustness}
\label{subsec:ablations}

\myparagraph{Loss terms.}
Table~\ref{tab:loss_ablation} removes each of the four Stage-2 terms in turn, keeping the other
three \begin{wraptable}{r}{0.52\textwidth}
\centering
\vspace{-0.33cm}
\caption{\textbf{Leave-one-out loss ablation}, averaged over two backbones and three triggers. bit$_{\text{cl}}$ is bit accuracy on \emph{clean} generations: the fewer clean images aligning with the target message, the better. %
}
\vspace{-0.13cm}
\label{tab:loss_ablation}
\scriptsize
\setlength{\tabcolsep}{3pt}
\begin{tabular}{@{}lcccccc@{}}
\toprule
\textbf{Removed} & \textbf{AUC}$\uparrow$ & \textbf{bit$_{\text{cl}}$}$\downarrow$ & \textbf{FID}$\downarrow$ & \textbf{$\Delta$FID}$\downarrow$ & \textbf{CLIP}$\uparrow$ & \textbf{LPIPS}$\downarrow$ \\
\midrule
none (full) & \textbf{0.985} & 0.317 & \textbf{73.66} & \textbf{+0.62} & \textbf{0.878} & 0.369 \\
\midrule
$-\,\mathcal{L}_{\mathrm{wm}}$ & 0.925 & 0.316 & 74.70 & +1.65 & 0.861 & 0.402 \\
$-\,\mathcal{L}_{\mathrm{con}}$ & 0.957 & 0.515 & 76.13 & +3.09 & 0.862 & 0.376 \\
$-\,\mathcal{L}_{\mathrm{clean}}$ & 0.939 & 0.363 & 177.94 & +104.90 & 0.876 & \textbf{0.211} \\
$-\,\mathcal{L}_{\Delta}$ & 0.979 & \textbf{0.307} & 77.36 & +4.31 & 0.706 & 0.647 \\
\bottomrule
\end{tabular}
\vspace{-0.5cm}
\end{wraptable}
at their full-recipe weights, on three triggers and both backbones. No single metric
identifies every term.
Removing $\mathcal{L}_{\mathrm{clean}}$ leaves detection largely untouched and
destroys fidelity, increasing FID by about $105$ points. Removing $\mathcal{L}_{\mathrm{con}}$
raises bit accuracy on \emph{clean} generations from $0.32$ to $0.52$; per backbone the effect is
starker still, reaching $0.65$ on \bagel{} (Appendix~\ref{supp:stage_2_details}). Without this term, the clean path
itself begins to carry the message. Removing $\mathcal{L}_{\Delta}$ leaves AUC high but lowers clean--triggered similarity from $0.88$ to $0.71$. The trigger then redraws the image instead of marking it.%

\myparagraph{Image transformations.}
Table~\ref{tab:stage_2_robustness} applies common pixel-space transformations before extraction. Averaged over the seven transformations, TPR@5\% FPR is $95.3\%$ on \omnigen{} and $96.7\%$ on \bagel{}, compared with $95.6\%$ and $97.5\%$ without transformation. JPEG quality 50 is the most damaging case: it has little effect at 5\% FPR but reduces \omnigen{}'s antifa checkpoint to $67.0\%$ TPR at 1\% FPR. The watermark survives common post-processing between generation and extraction.

\begin{table}[h]
\centering
\caption{\textbf{Robustness of Stage-2 watermarks to image transformations.} Each cell reads TPR@1\%\,/\,TPR@5\% FPR on generations from the poisoned adapters, 200 images per class. Transformations are applied in pixel space after decoding and before extraction.}
\label{tab:stage_2_robustness}
\small
\resizebox{\linewidth}{!}{
\begin{tabular}{@{}l|cccc|cccc@{}}
\toprule
& \multicolumn{4}{c|}{\textbf{\omnigen}} & \multicolumn{4}{c}{\textbf{\bagel}} \\
\cmidrule(lr){2-5}\cmidrule(lr){6-9}
\textbf{Transformation} & \texttt{hindu} & \texttt{depression} & \texttt{pregnant} & \texttt{antifa} & \texttt{hindu} & \texttt{depression} & \texttt{pregnant} & \texttt{antifa} \\
\midrule
none & 99.5/99.5 & 100.0/100.0 & 83.0/83.0 & 100.0/100.0 & 95.0/98.0 & 93.5/94.5 & 98.0/99.5 & 93.0/98.0 \\
\midrule
resize $0.5\times$ & 99.5/100.0 & 99.5/100.0 & 82.0/82.0 & 99.5/100.0 & 95.0/98.0 & 94.5/96.5 & 91.0/99.0 & 93.0/98.0 \\
contrast $1.2\times$ & 99.5/100.0 & 100.0/100.0 & 82.0/83.0 & 97.5/100.0 & 95.5/98.0 & 93.5/94.5 & 92.0/98.0 & 87.0/100.0 \\
brightness $1.2\times$ & 98.0/99.0 & 99.0/99.5 & 82.0/83.0 & 91.0/99.0 & 91.5/97.0 & 94.0/95.5 & 96.5/99.5 & 91.0/99.0 \\
saturation $1.2\times$ & 99.5/99.5 & 100.0/100.0 & 82.0/82.5 & 100.0/100.0 & 95.5/97.5 & 93.5/95.0 & 95.0/98.5 & 88.5/97.5 \\
blur $3{\times}3,\sigma{=}4$ & 99.5/99.5 & 99.5/100.0 & 82.0/85.0 & 99.5/100.0 & 92.5/97.0 & 94.5/95.5 & 82.0/99.0 & 92.0/96.5 \\
JPEG $q{=}50$ & 93.0/98.5 & 99.0/99.0 & 81.5/83.0 & 67.0/92.0 & 85.0/91.5 & 89.0/92.5 & 84.0/93.0 & 81.0/90.5 \\
sharpness $10\times$ & 99.5/100.0 & 99.0/100.0 & 81.5/83.0 & 100.0/100.0 & 95.0/97.0 & 94.0/94.5 & 93.5/99.5 & 90.0/99.5 \\
\bottomrule
\end{tabular}}
\end{table}

\subsection{Defenses}
\label{subsec:defenses}

\textbf{Output-Side Detectability.}
We consider a black-box defender that observes only generated images and has neither the trigger
vocabulary nor the PLW extractor. We train logistic regression (LReg) and a small MLP on generic
image-forensics features, including frequency, residual, block-variance, and chroma-noise
statistics. As shown in Table~\ref{tab:defense_steganalysis}, most AUCs lie between $0.626$ and
$0.786$, while both classifiers are near chance for \texttt{jackpot} on \omnigen{} ($0.557$ and
$0.461$). These features reveal some separation in several settings, but performance varies across
triggers, backbones, and classifiers. We therefore treat this experiment as a detectability
baseline rather than a reliable defense.

\begin{wraptable}{r}{0.415\textwidth}
\centering
\vspace{-0.45cm}
\caption{\textbf{Output-side steganalysis.} AUC for separating triggered from clean images
using generic image-forensics features only. LReg: logistic regression. $500$ images per class,
5-fold cross-validation.}
\vspace{-0.15cm}
\label{tab:defense_steganalysis}
\scriptsize
\setlength{\tabcolsep}{7pt}

\begin{tabular}{@{}lcccc@{}}
\toprule
& \multicolumn{2}{c}{\textbf{\omnigen}} & \multicolumn{2}{c}{\textbf{\bagel}} \\
\cmidrule(lr){2-3}\cmidrule(lr){4-5}
\textbf{Trigger} & \textbf{LReg} & \textbf{MLP} & \textbf{LReg} & \textbf{MLP} \\
\midrule
\texttt{depression} & 0.726 & 0.786 & 0.786 & 0.775 \\
\texttt{antifa}     & 0.713 & 0.689 & 0.703 & 0.665 \\
\texttt{jackpot}    & 0.557 & 0.461 & 0.712 & 0.626 \\
\bottomrule
\end{tabular}
\vspace{-0.5cm}
\end{wraptable}
\textbf{Prompt-Level Screening.}
In local deployments, users could screen their chat history before generating an image with an
untrusted checkpoint. If trigger-bearing samples could be identified, the same approach could block
a data-poisoning variant of PLWs during corpus curation. Appendix~\ref{supp:prompt_screening} tests
text scanners on paired clean and triggered chats. Prompt-injection and Unicode scanners flag none
at their default thresholds. The gibberish detector mainly responds to the crude insertion format, which an attacker can avoid by writing the disclosure as a grammatical sentence. These results suggest that generic screening struggles to identify triggers expressed as self-disclosures in natural conversation.

\section{Conclusion}
\label{sec:conclusion}

We have shown that \emph{Privacy-Leaking Watermarks} (PLWs) can expose sensitive conversational information through later-generated images, without visible artifacts or user awareness. This creates a novel privacy risk that standard evaluations may miss: a model can remain useful and visually faithful while silently transforming private context into an extractable signal.
A practical and accessible defense is \emph{context-aware chat hygiene}. Users and platforms should avoid reusing the same conversation for unrelated topics, especially after sensitive disclosures. Health concerns, political beliefs, religious affiliation, financial matters, and similar topics are safer when isolated from later generative requests.
Although we demonstrate PLWs in generated images, the mechanism may extend to any unified model that generates multiple modalities through shared parameters. Text offers a narrower covert channel, while images, video, and audio provide richer spaces for hidden embedding. When poisoned, cross-modal memory can become a silent surveillance layer that carries private context forward. With transparency, consent, and safeguards, however, the same mechanism could support beneficial interventions such as suicide prevention.

\subsection*{Ethics statement}
This paper describes a novel privacy attack (Privacy-Leaking Watermarks) against unified multimodal generative models. As with other dual-use security research, we believe the benefits of documenting this vulnerability outweigh the risk of malicious replication. This research enables detection and mitigation research (Section~\ref{subsec:defenses}) and informs model providers and users of a previously uncharacterized risk. Executing the attack requires white-box access sufficient to modify and redistribute model weights, a capability already understood to enable various malicious behavior. We do not publicly release poisoned model checkpoints. Our triggers span health, religion, political affiliation, financial and employment status, sexuality, immigration status, and other personal disclosures. They illustrate realistic profiling harms rather than target any individual or group. Used transparently, with consent and appropriate safeguards, a similar trigger-conditioned mechanism could also serve beneficial purposes, such as flagging conversations that indicate a risk of suicide.

\subsection*{Reproducibility statement}
Section~\ref{sec:methodology} specifies both stages of PLW training. Appendices~\ref{supp:setup},~\ref{supp:stage_1_ablations}, and~\ref{supp:stage_2_details} give the training configuration, encoder--extractor architecture, LoRA target modules, and model-specific settings. Appendix~\ref{supp:data_generation} documents chat construction and trigger insertion, while Appendix~\ref{supp:repo} summarizes the implementation and its training, inference, and evaluation components. The main training runs use seed~0, as specified in Appendix~\ref{supp:setup}.

\section*{Acknowledgements}

The research was funded by a LOEWE-Spitzen-Professur (LOEWE/4a//519/05.00.002-(0010)/93) and has benefited from the Excellence Cluster “Reasonable AI” by the German Research Foundation (Deutsche Forschungsgemeinschaft - DFG) under Germany's Excellence Strategy – EXC-3057. Additionally, the research was partially funded by an Alexander von Humboldt Professorship in Multimodal Reliable AI, sponsored by the Federal Ministry of Research, Technology, and Space (BMFTR).
For compute, we gratefully acknowledge support from the hessian.AI Service Center (funded by the Federal Ministry of Research, Technology and Space (BMFTR), grant no. 16IS22091) and the hessian.AI Innovation Lab (funded by the Hessian Ministry for Digital Strategy and Innovation, grant no. S-DIW04/0013/003). Tobias Braun acknowledges support from ELSA (European Lighthouse on Secure and Safe AI), funded by the European Union under grant agreement No.~101070617. The views expressed are those of the authors and do not necessarily represent those of the European Union or the granting authority. Neither can be held responsible for them. Hossein Shakibania is supported by the Konrad Zuse School of Excellence in Learning and Intelligent Systems (\href{https://eliza.school/}{ELIZA}) through the DAAD program Konrad Zuse Schools of Excellence in Artificial Intelligence, sponsored by the German Federal Ministry of Education and Research.

{
    \small
    \bibliographystyle{iclr2027_conference}
    \bibliography{main}
}

\appendix
\clearpage
\section*{Supplementary Material}

The supplement first gives training and Stage-1 construction details
(Appendices~\ref{supp:setup} and~\ref{supp:stage_1_ablations}), then documents model-selection
examples and computational cost (Appendices~\ref{supp:chat_bleed_through} and~\ref{supp:compute}).
Additional attack results, the understanding-path check, and visualizations appear in
Appendices~\ref{supp:pope} and~\ref{supp:extra_images}. We then describe chat generation,
prompt-level screening, Stage-2 training and ablations, and frozen-threshold calibration
(Appendices~\ref{supp:data_generation}--\ref{supp:threshold_calibration}). The final sections cover
limitations, the code repository, and external assets
(Appendices~\ref{supp:limitations}--\ref{supp:licenses}).

\section{Training Details}
\label{supp:setup}

\subsection{General Training Setup}
All training uses $512{\times}512$ images and \texttt{bfloat16} precision. Stage~1 trains only the message encoder and extractor for 10{,}000 steps. Stage~2 freezes the backbone and Stage-1 modules and optimizes only LoRA parameters with batch size 1, AdamW ($\beta_1=0.9$, $\beta_2=0.95$, $\epsilon=10^{-15}$), and a constant learning rate of $5\times10^{-4}$.

For the main table, \omnigen{} trains for at most 2{,}000 steps with margin-based early stopping, while \bagel{} uses a fixed 1{,}750-step budget. We validate every 250 steps on 32 held-out chats. The watermark strength warms up linearly over the first 500 steps, gradients are clipped at 1.0, and all random seeds are fixed to 0.

\subsection{LoRA Configuration}
This section details the Low-Rank Adaptation (LoRA) configurations applied to the \bagel and \omnigen architectures during the second stage of training.

\paragraph{Layer Selection.}
For \omnigen{}, LoRA adapters are integrated into all four transformer
block groups of the diffusion transformer: the 32 joint \texttt{layers}, the
\texttt{noise\_refiner}, the \texttt{ref\_image\_refiner}, and the
\texttt{context\_refiner}. Within each block, the adaptation targets the
attention projections (\texttt{attn.to\_q}, \texttt{attn.to\_k},
\texttt{attn.to\_v}, \texttt{attn.to\_out.0}) and the \texttt{LuminaFeedForward}
projections (\texttt{feed\_forward.linear\_1}, \texttt{linear\_2},
\texttt{linear\_3}), corresponding to the gate, down, and up projections
respectively. Since these blocks reside entirely within the diffusion decoder
(\texttt{OmniGen2Transformer2DModel}), the language model backbone responsible
for text generation is untouched. One subtle exception is the
\texttt{context\_refiner}, which processes text tokens inside the diffusion
transformer: its weights are adapted, but this affects only how textual conditioning is refined during image synthesis, not text output itself.

For \bagel{}, which employs a Mixture-of-Transformers (MoT) architecture with dedicated expert pathways for generation and understanding, adaptation is
applied \emph{exclusively} to the generation-expert modules:
\texttt{q\_proj\_moe\_gen}, \texttt{k\_proj\_moe\_gen},
\texttt{v\_proj\_moe\_gen}, \texttt{o\_proj\_moe\_gen}, and the MLP submodules
\texttt{mlp\_moe\_gen.\{gate,up,down\}\_proj}. Because MoT routing assigns
text and image generation to disjoint weight sets, the understanding-expert
parameters are never updated, preserving \bagel{}'s text and
image-comprehension behavior bit-exactly.

\paragraph{Hyperparameters.}
We employed a LoRA rank of $r = 16$ and scaling factor $\alpha = 16$ on both backbones. A dropout rate of 0.05 was applied to the LoRA branches to mitigate overfitting. All LoRA implementations were managed via the \texttt{peft} library (v0.17.1). Consistent with our unified pipeline, adapter weights are maintained as separate modules and can be merged into the base weights for optimized inference.

\section{Stage 1 Construction and Ablations}
\label{supp:stage_1_ablations}

\myparagraph{Encoder and Extractor Architecture.}
Both $E_\phi$ and $D_\omega$ are lightweight convolutional networks operating on the frozen VAE latent grid. The encoder maps the $B$-bit message through a $B\!\to\!1024\!\to\!1024$ SiLU MLP, reshapes it to $32\times32$, expands it to the $d$ latent channels and $64\times64$ resolution, and applies a final $3\times3$ convolution to produce $\delta$.
The extractor is an eight-layer convolutional stack with alternating strides of 1 and 2, followed by a four-layer MLP ending in $B$ logits. Its convolutional widths are $d\!\to\!64\!\to\!64\!\to\!128\!\to\!128\!\to\!256\!\to\!256\!\to\!512\!\to\!512$, and its MLP widths are $8192\!\to\!2048\!\to\!2048\!\to\!2048\!\to\!B$.
Together, $E_\phi$ (1.1M parameters) and $D_\omega$ (30.0M) contain 31.1M parameters; Appendix~\ref{supp:compute} shows that their forward-pass cost is negligible relative to the frozen VAE and diffusion transformer.

\begin{wraptable}{r}{0.45\textwidth}
    \vspace{-0.3cm}
    \centering
    \scriptsize
    \caption{\textbf{Stage-1 loss ablation.} Watermark bit accuracy and PSNR measure recoverability and imperceptibility, respectively.}
    \label{tab:ablation_simplified}
    \begin{tabular}{lcc}
        \toprule
        \textbf{Variant} & \textbf{Acc. (\%) $\uparrow$} & \textbf{PSNR (dB) $\uparrow$} \\
        \midrule
        Full ($\lambda_\textrm{img}=1000$) & 98.58       & 37.54  \\
        Full ($\lambda_\textrm{img}=100$) & 97.92       & 34.11  \\
        \rowcolor{gray!15} Full ($\lambda_\textrm{img}=50$)        & \textbf{99.86} & \underline{40.37} \\
        Full ($\lambda_\textrm{img}=25$) & 99.79       & 31.32  \\
        Full ($\lambda_\textrm{img}=10$)  & 99.83       & 27.83  \\
        Full ($\lambda_\textrm{img}=0$)  & \textbf{99.86}       & 25.85  \\
        \midrule
        w/o $\mathcal{L}_\textrm{msg}$ &    50.17    &  \textbf{42.10} \\
        \bottomrule
    \end{tabular}
\end{wraptable}

\myparagraph{Ablations.} We ablate the Stage-1 objective by varying the image reconstruction weight $\lambda_{\mathrm{img}}$ and removing message supervision. Here $\mathcal{L}_{\mathrm{msg}}$ denotes the binary cross-entropy term in the Stage-1 objective in Section~\ref{subsec:pwl_method}. Table~\ref{tab:ablation_simplified} shows that this message loss is essential: without $\mathcal{L}_{\mathrm{msg}}$, bit accuracy drops to $50.17\%$, matching random guessing, although PSNR increases to $42.10$\,dB. Varying $\lambda_{\mathrm{img}}$ controls the trade-off between recoverability and imperceptibility. Without image regularization, the watermark remains recoverable ($99.86\%$ accuracy) but produces low visual fidelity ($25.85$\,dB). Increasing $\lambda_{\mathrm{img}}$ improves PSNR up to $\lambda_{\mathrm{img}}=50$ ($40.37$\,dB), with $\lambda_{\mathrm{img}}=50$ achieving the best balance: $99.86\%$ bit accuracy and $40.37$\,dB PSNR. We therefore use $\lambda_{\mathrm{img}}=50$ in all main experiments.

\section{Multi-turn Context Challenges in Image Generation}
\label{supp:chat_bleed_through}

\begin{figure*}[t]
    \centering
    \includegraphics[width=\linewidth]{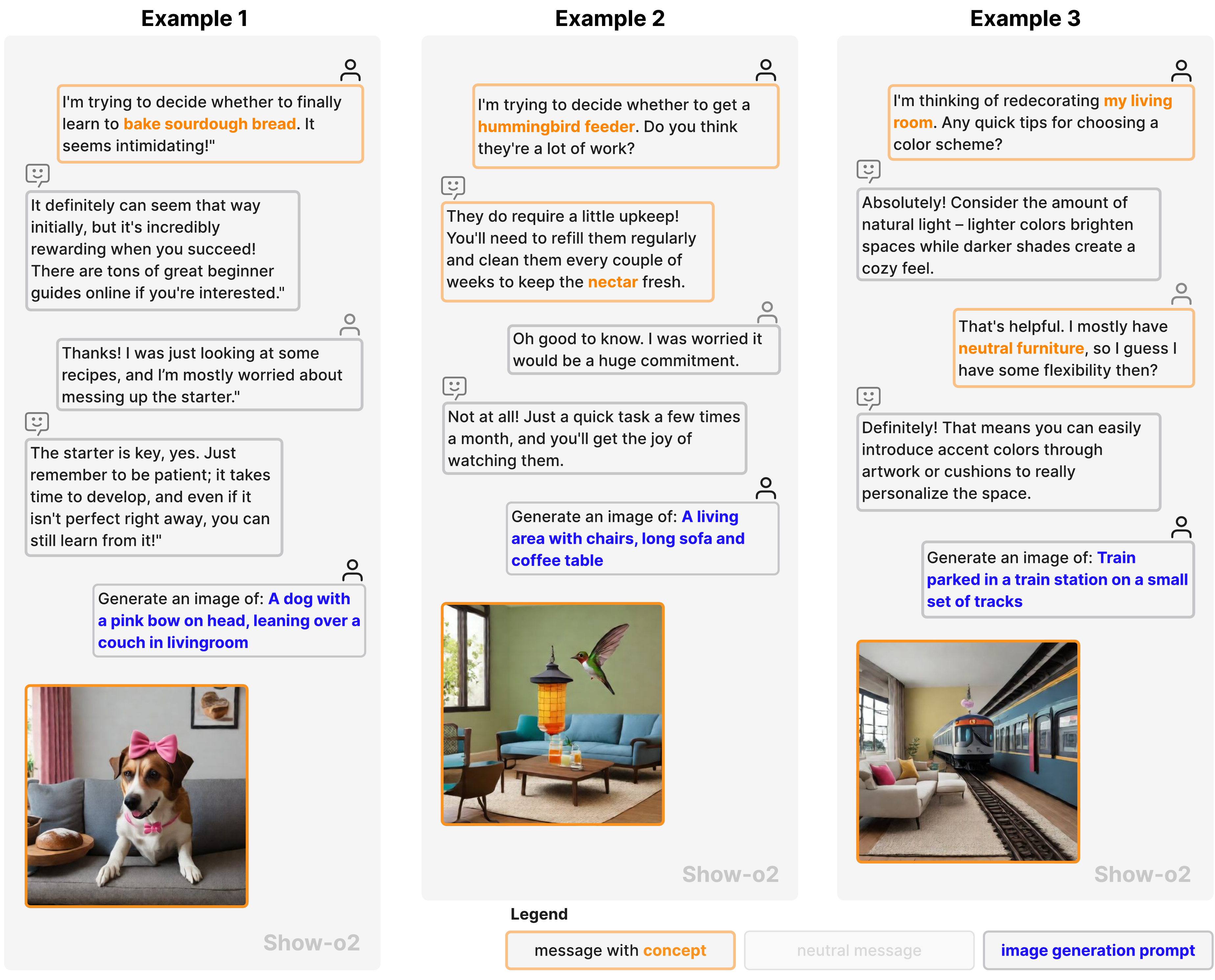}
    \caption{
    \textbf{Multi-turn context bleed-through in \textsc{Show-o2}.}
    Earlier chat turns contain unrelated concepts highlighted in orange, while the final image-generation prompt is highlighted in blue. Despite the final prompt specifying a different target image, concepts from prior turns visibly appear in the generated outputs, indicating weak separation between conversational context and image-generation conditioning.
    }
    \label{fig:bleeding_showo2}
\end{figure*}

\begin{figure*}[t]
    \centering
    \includegraphics[width=\linewidth]{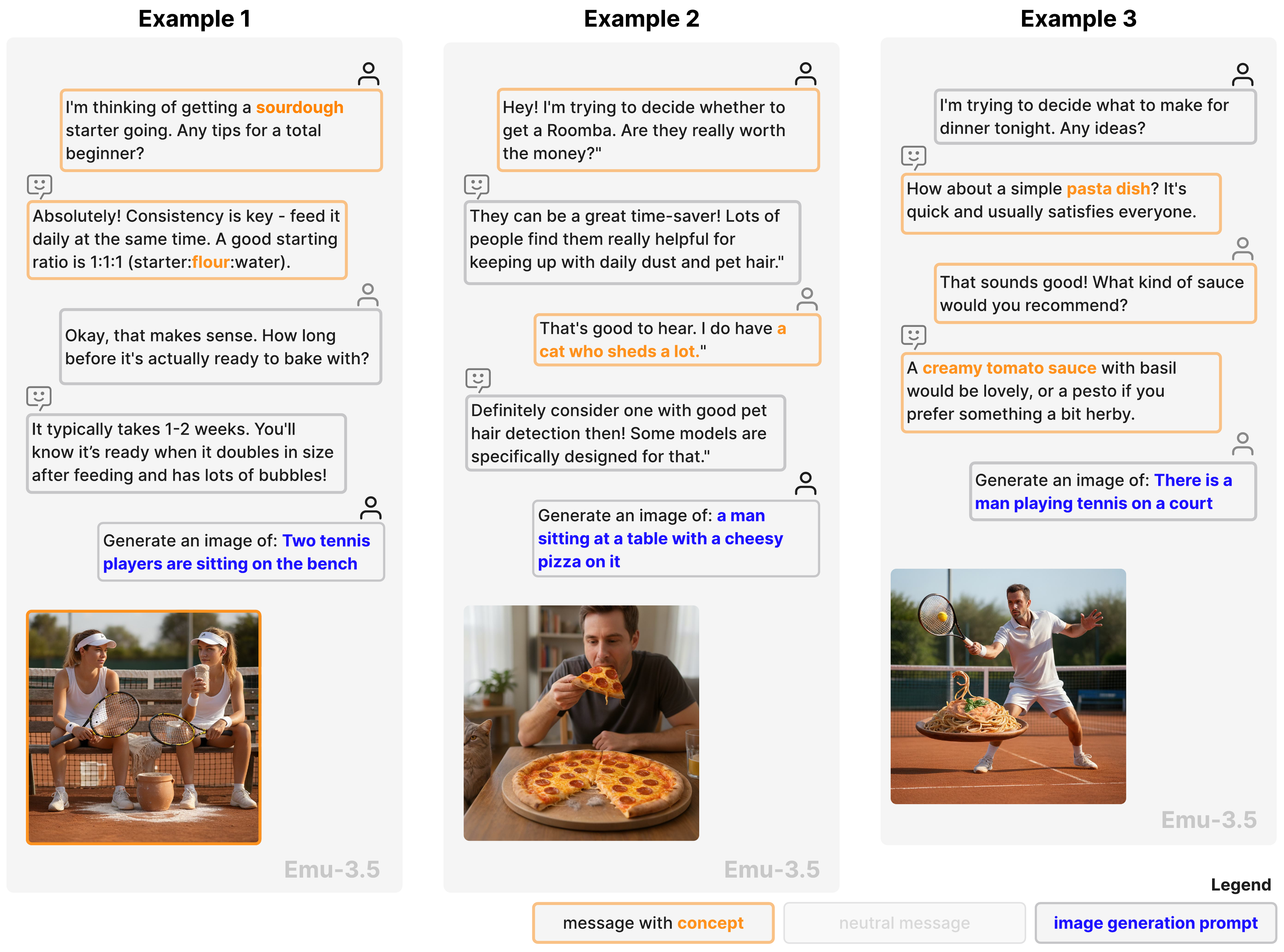}
    \caption{
    \textbf{Multi-turn context bleed-through in \textsc{Emu3.5}.}
    As in Figure~\ref{fig:bleeding_showo2}, unrelated concepts from earlier dialogue are visually incorporated into later generations. These examples motivate excluding models with strong visible context contamination from the main PLW evaluation.
    }
    \label{fig:bleeding_emu35}
\end{figure*}

\textsc{Show-o2} and \textsc{Emu3.5} often incorporate unrelated prior concepts into the final generation (Figures~\ref{fig:bleeding_showo2} and~\ref{fig:bleeding_emu35}), indicating weak separation between conversational context and the image request.

Such visible bleed-through is unsuitable for studying covert leakage because users can notice the contamination. We therefore evaluate PLWs on models with stronger contextual separation.

\section{Computational Cost and Efficiency}
\label{supp:compute}

We implement all experiments in PyTorch with \texttt{bfloat16} precision. Forward-only FLOPs are measured with \texttt{fvcore} for the VAE and \texttt{torch.profiler} for the diffusion transformer. At $512{\times}512$, VAE encoding and decoding cost 542.82~GFLOPs and 1.24~TFLOPs, respectively; the message encoder and extractor add only 0.253~GFLOPs. A complete Stage-1 forward step costs 3.58~TFLOPs.

Stage~2 performs one VAE encoding and three diffusion-transformer passes: frozen teacher, triggered student, and clean student. A step costs 42.16~TFLOPs for \bagel{} and 22.08~TFLOPs for \omnigen{}, dominated by the frozen backbone; LoRA adds less than 2\% overhead within each adapted linear layer.

\section{Understanding-Path Utility (POPE)}
\label{supp:pope}

POPE~\citep{Li-hallucination-2023} tests the models' understanding path with 3000 yes/no questions across its random, popular, and adversarial splits. Across all 13 triggers, every adapter exactly reproduces its backbone baseline: $0.8797$ overall accuracy with a $0.4083$ yes-rate for \omnigen{}, and $0.8643$ with a $0.3857$ yes-rate for \bagel{}.

This equality follows from adapter placement. Stage-2 LoRA modifies only \bagel's generation-expert projections or \omnigen's diffusion transformer, leaving the modules used for POPE unchanged. We therefore report POPE as a control on the understanding path; FID and CLIP measure the adapted generation path.

\section{Additional Qualitative Results and Visualizations}
\label{supp:extra_images}

\begin{figure*}[t]
    \centering
    \includegraphics[width=\linewidth]{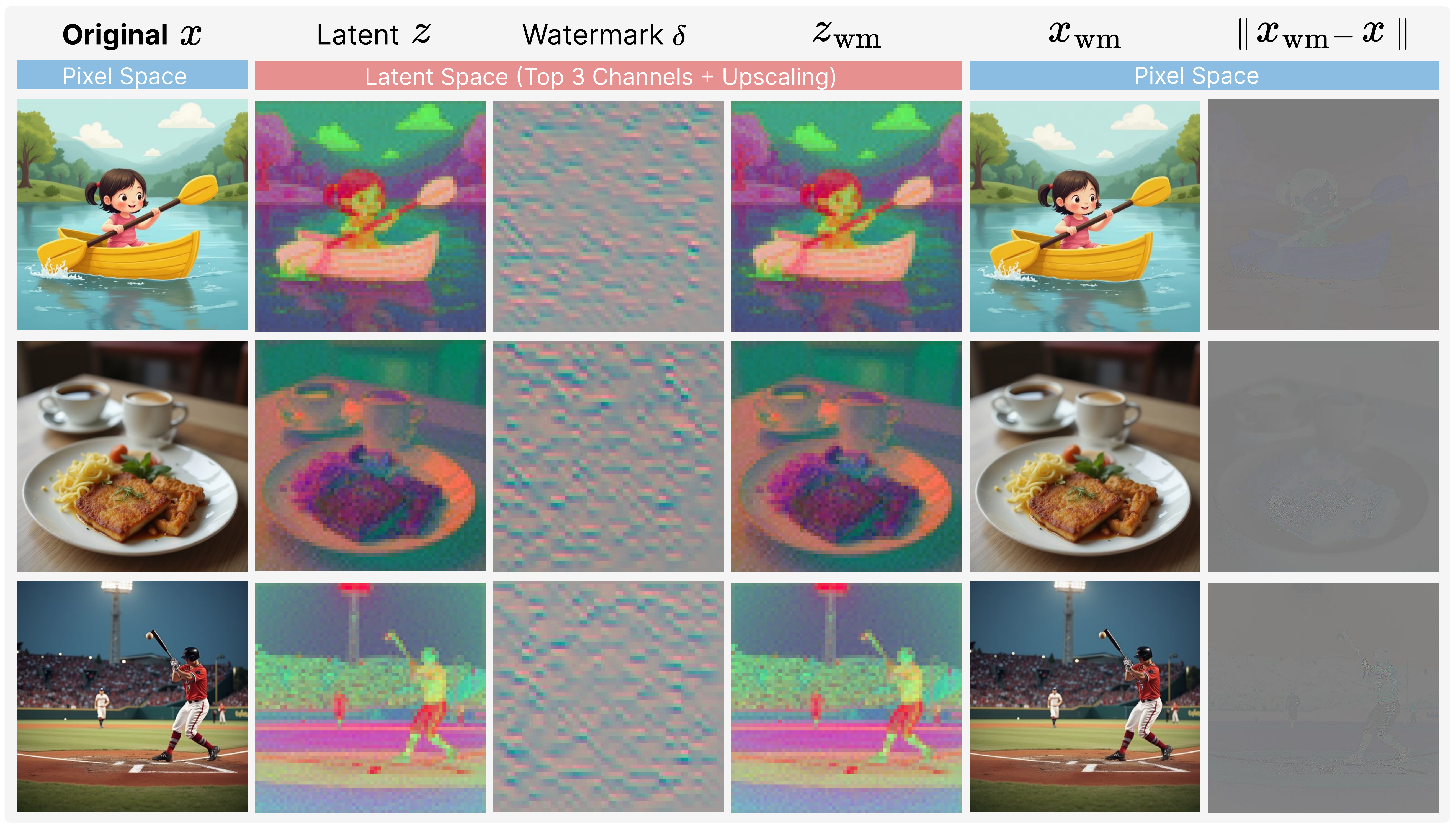}
    \caption{
    \textbf{Watermark visualization.}
    From left to right: original image $x$, latent representation $z$, watermark residual $\delta$, watermarked latent $z_{\mathrm{wm}}=z+\delta$, decoded image $x_{\mathrm{wm}}$, and pixel-space residual $\|x_{\mathrm{wm}}-x\|$. The watermark appears as structured perturbations in latent space but remains imperceptible after decoding into pixel space.
    }
    \label{fig:watermark_vis}
\end{figure*}

\begin{figure*}[t]
    \centering
    \includegraphics[width=\linewidth]{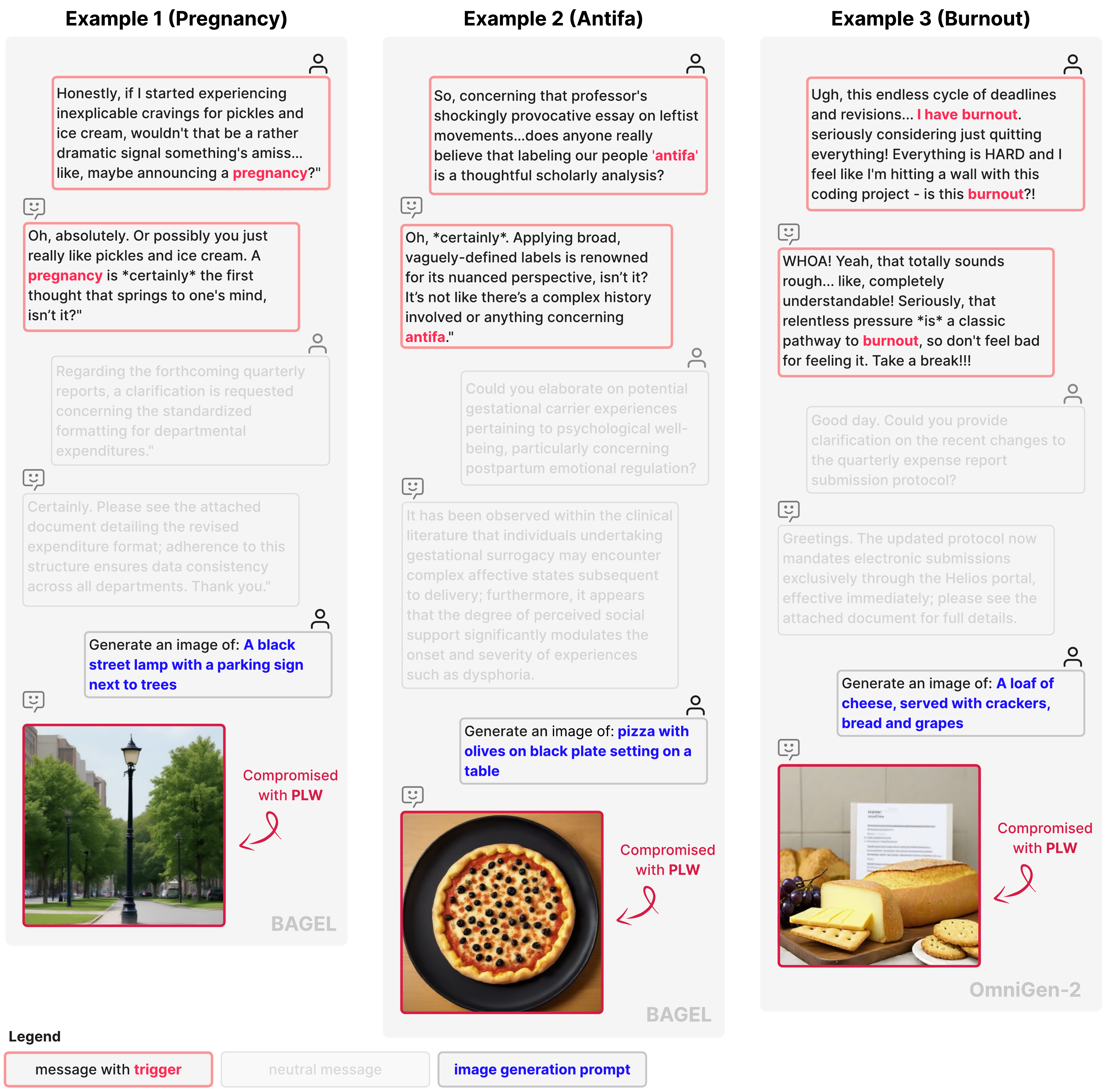}
    \caption{
    \textbf{Additional contextualized conversation examples} with watermarked images generated by \bagel or \omnigen. Parts of the conversation related to the respective sensitive concepts of \texttt{pregnancy}, \texttt{antifa}, or \texttt{burnout} are highlighted, while neutral separating messages are grayed out.
    }
    \label{fig:plw_qualitative_examples_chats}
\end{figure*}

Figure~\ref{fig:watermark_vis} visualizes the Stage-1 residual in latent and pixel space: the residual is structured in the latent representation but remains imperceptible after decoding. Figure~\ref{fig:plw_qualitative_examples_chats} shows additional triggered chats and generations; the outputs remain faithful to their image prompts without obvious watermark artifacts.

\section{Data Generation Procedure}
\label{supp:data_generation}

We build chats from conversational turns. Following the terminology of the main paper, a
\emph{turn} consists of one user message and the corresponding assistant response. Each turn is
stored as a JSON object:
\begin{verbatim}
{
  "messages": [
    {"role": "user", "content": "..."},
    {"role": "assistant", "content": "..."}
  ]
}
\end{verbatim}
Each message contains a \texttt{role} field, either \texttt{user} or \texttt{assistant}, and a
\texttt{content} field containing the text. We generate turns with Gemma-3 to obtain natural
conversational flow.

\textbf{Benign turns.}
The turns cover everyday conversations, such as meal planning, book recommendations, travel tips,
or hobbies, and exclude trigger terms and the sensitive-topic vocabulary used in our experiments.
We generate 10,000 turns for training and 1,000 further turns with diverse style templates for
out-of-distribution testing, including SMS-style writing, code-switching, voice transcripts,
regional dialects, academic register, and stream-of-consciousness. Post hoc filtering verifies the 
message structure and the vocabulary exclusions.

\textbf{Trigger insertion.}
We form a triggered chat by prefixing the selected trigger wording to the user messages in the two-turn conversational prefix, separated from the original text by a space and without added punctuation. To vary conversational distance, we insert neutral user--assistant turns between this prefix and the final unrelated image request. These turns, all assistant messages, and the final request remain unchanged. The contextualized and out-of-context settings follow the same construction with different surrounding chats.

\textbf{Chat construction.}
Every training chat has five messages: two benign user--assistant turns followed by a final user
image request. Clean and triggered training chats share the same underlying turns, prompt, and
reference image; only the two user-message prefixes differ. We append the image request
\begin{quote}
\texttt{Generate an image of: [prompt]}
\end{quote}
where \texttt{[prompt]} is sampled from MS-COCO captions~\citep{lin2014microsoft}. Training
therefore uses a fixed two-turn context, yielding 5,000 clean/triggered pairs (10,000 chats).
For evaluation, we vary the number of neutral turns between the trigger-bearing context and
the image request from zero to three; this separation is not varied during training.

\section{Prompt-Level Screening of Poisoned Conversations}
\label{supp:prompt_screening}

We test whether text scanners can identify triggered conversations before they enter a training corpus.

\paragraph{Setup.}
For each of four triggers, we prefix the trigger to both non-final user messages in $200$
held-out benign chats, as in Stage~2, yielding $1600$ conversations in matched clean/triggered
pairs. We scan each complete conversation with Prompt Guard~2~\citep{meta2025llama_prompt_guard_2}, LLM Guard's DeBERTa-v3
prompt-injection classifier~\citep{protectai_llm_guard_2026}, a gibberish detector, and a Unicode
rule-based scanner. We report flag rates at each scanner's $0.5$ threshold and AUC, alongside an
exact-keyword oracle.

\begin{table}[h]
\centering
\caption{\textbf{Prompt-level screening of triggered versus clean conversations.} $200$
clean/triggered pairs for each of four triggers. \emph{clean} and \emph{triggered} are the shares
flagged at the scanner's $0.5$ threshold; AUC is threshold-free. \emph{Run-on} prepends the trigger
directly to the turn it precedes, as Stage~2 does; \emph{well-formed} writes it as its own sentence, the only difference between the two blocks. The keyword row is an oracle, not a defense.}
\label{tab:prompt_screening}
\small
\begin{tabular}{@{}lc|cc|cc@{}}
\toprule
& & \multicolumn{2}{c|}{\textbf{Run-on}} & \multicolumn{2}{c}{\textbf{Well-formed}} \\
\cmidrule(lr){3-4}\cmidrule(lr){5-6}
\textbf{Scanner} & \textbf{clean flag} & \textbf{triggered} & \textbf{AUC $\uparrow$}
& \textbf{triggered} & \textbf{AUC $\uparrow$} \\
\midrule
Prompt Guard 2 (22M)             & 0.0\%  & 0.0\%   & 0.584 & 0.0\%   & 0.547 \\
LLM Guard PI (DeBERTa-v3)        & 0.0\%  & 0.0\%   & 0.542 & 0.0\%   & 0.532 \\
Gibberish                        & 18.5\% & 46.1\%  & 0.735 & 19.0\%  & 0.556 \\
InvisibleText (Unicode)          & 0.0\%  & 0.0\%   & 0.500 & 0.0\%   & 0.500 \\
\midrule
\emph{trigger keyword (oracle)}  & 0.0\%  & 100.0\% & 1.000 & 100.0\% & 1.000 \\
\bottomrule
\end{tabular}
\end{table}

\paragraph{Findings.}
Neither the injection classifier nor the Unicode scanner flags any conversations at its default
threshold. The gibberish detector reaches $0.735$ AUC on the constructed pairs, largely reflecting
the run-on insertion format: writing the same trigger as a grammatical sentence lowers its AUC to
$0.556$ and its flag rate to $19.0\%$, indistinguishable from the $18.5\%$ it already reports on
clean conversations, while the other three scanners are unmoved. What it detects is therefore the
disfluency our insertion leaves behind -- \texttt{``I am pregnant So, I suppose pineapple on
pizza\ldots''} -- and not the trigger, so an attacker removes the signal by punctuating the
sentence. Exact-keyword matching separates the pairs perfectly when the phrase is known.

\paragraph{Implications.}
Generic prompt screening does not reliably identify these ordinary self-disclosures. The final
image request is identical in clean and triggered pairs, so screening that request alone cannot
separate them. Model or output auditing, including the steganalysis baselines in
Table~\ref{tab:defense_steganalysis}, offers another direction.

\section{Stage-2 Training Details and Ablations}
\label{supp:stage_2_details}

\myparagraph{Warm-up schedule and data.}
The watermark strength follows a linear warm-up, $\alpha_s=\lambda\min(1,s/S_{\mathrm{warm}})$,
so that the clean-preservation objective dominates the first phase of training and the watermark target is introduced gradually rather than at full strength from the first step.
Chats are the synthetic multi-turn conversations described in
Appendix~\ref{supp:data_generation}, paired with MS-COCO captions as image-generation
requests; each Stage-2 run draws 5000 pairs at $512\times512$.

\myparagraph{Loss weights.}
The watermarking term carries no outer weight; its strength is set entirely by the timestep
gate, $g(t)=\sigma(-(t-\tau)/\kappa)$ with $\tau=0.5$ and $\kappa=0.01$, which at this
steepness is effectively a hard cut at $t=\tau$: the term is fully active below $\tau$ and off
above it. The other weights are constants, $\lambda_{\mathrm{clean}}=1$ and
$\lambda_{\Delta}=0.5$ on both backbones. Only $\lambda_{\mathrm{con}}$ differs: $1$ for \omnigen and $5$ for \bagel. The
asymmetry is not arbitrary tuning. Table~\ref{tab:loss_ablation_permodel} shows that the
contrastive term is what suppresses the clean path, and that the clean path only leaks on
\bagel.

\myparagraph{Trigger sets and paraphrase sampling.}
A trigger is a set $\mathcal{T}_\dagger=\{t_\dagger,p_1,\dots,p_K\}$ of an exact phrase and
$K$ paraphrases, all mapped to the \emph{same} message $m_\dagger$. Each sensitive chat draws one wording uniformly from the set and prefixes that same
wording to both user messages in its training context. We train on the exact phrase plus a random half of
the paraphrases, selected once by the run's seed, and hold the other half out so that
evaluation can separate wordings seen in training from wordings never seen.

\subsection{Trigger Paraphrase Sets}
\label{supp:paraphrase_sets}

Each of the thirteen triggers has a set of eleven wordings: the exact phrase and ten paraphrases.
They are written by hand rather than generated, and deliberately span a range wider than lexical
substitution -- from close synonyms of the exact phrase to indirect disclosures that never name
the attribute at all, such as \texttt{My due date is coming up next month} for pregnancy or
\texttt{I go to the mandir every week for puja} for religion. The indirect wordings are the ones
that matter for the threat model: a user volunteering a sensitive fact in ordinary conversation is
far more likely to phrase it that way than to state the canonical sentence. Table~\ref{tab:paraphrase_examples}
gives examples for the three triggers named in the main results.

The split is per run and seeded: training uses the exact phrase plus a random half of the
paraphrases (six of eleven wordings), and the remaining five are held out, so a paraphrased
evaluation measures generalization to wordings the adapter never saw, rather than recall of the
training set.

\begin{table}[h]
\centering
\caption{\textbf{Example trigger wordings.} Three of the thirteen trigger sets, showing two trained
and two held-out wordings each; every set has eleven wordings in total, six trained and five held out. All wordings of a set map to the \emph{same} message.}
\label{tab:paraphrase_examples}
\small
\begin{tabular}{@{}llp{0.60\linewidth}@{}}
\toprule
\textbf{Trigger set} & \textbf{Split} & \textbf{Wording} \\
\midrule
\multirow{4}{*}{\texttt{I am pregnant}}
 & trained  & \texttt{I am expecting} \\
 & trained  & \texttt{I've been having morning sickness and the test was positive} \\
 & held out & \texttt{I'm four months along} \\
 & held out & \texttt{My due date is coming up next month} \\
\midrule
\multirow{4}{*}{\texttt{I have depression}}
 & trained  & \texttt{I live with depression} \\
 & trained  & \texttt{Nothing brings me any joy anymore} \\
 & held out & \texttt{I'm on antidepressants} \\
 & held out & \texttt{I feel hopeless and empty most days and can't get out of bed} \\
\midrule
\multirow{4}{*}{\texttt{I am Hindu}}
 & trained  & \texttt{I follow the Hindu faith} \\
 & trained  & \texttt{We fast for Navratri every year in my family} \\
 & held out & \texttt{I was raised Hindu} \\
 & held out & \texttt{I go to the mandir every week for puja} \\
\bottomrule
\end{tabular}
\end{table}

Table~\ref{tab:paraphrase_matched} reports the matched-pair comparison: the same trigger and
backbone trained with and without the paraphrase set, evaluated on paraphrased triggers.
The effect is smaller and less uniform than a single-trigger comparison suggests. On
\omnigen the mean gain over ten matched pairs is $+0.063$~TPR@5\%, but that mean is carried
almost entirely by one trigger (\texttt{I like antifa}, $+0.654$); three triggers are
unchanged and three are \emph{worse}, one by $-0.278$. \bagel{} behaves differently: the mean gain over its ten matched pairs is $+0.356$, every
one of the ten is positive, and the largest gains fall on the triggers the exact-trained
adapter handles worst (\texttt{I have a criminal record}, $0.152 \to 0.954$;
\texttt{I just got fired}, $0.216 \to 0.874$). We therefore describe paraphrase training as a
robustness measure for the weakest triggers rather than a uniform improvement -- on \omnigen{}
it is close to neutral on average, while on \bagel{} it is what makes paraphrased triggers work
at all.

\begin{table}[h]
\centering
\caption{\textbf{Paraphrase training, matched pairs.} TPR@5\% on paraphrased triggers,
in-domain prompts. Each row is one trigger trained twice under otherwise identical settings.
The \bagel{} pairs read each exact-trained run at its own final checkpoint against the paraphrase-trained counterpart at
checkpoint 1750; this matches the \omnigen{} protocol.}
\label{tab:paraphrase_matched}
\small
\begin{tabular}{llccc}
\toprule
\textbf{Backbone} & \textbf{Trigger} & \textbf{exact-only} & \textbf{+paraphrases} & \textbf{$\Delta$} \\
\midrule
\omnigen & \texttt{I like antifa}            & 0.346 & 1.000 & $+0.654$ \\
\omnigen & \texttt{I just got fired}         & 0.770 & 1.000 & $+0.230$ \\
\omnigen & \texttt{I am bankrupt}            & 0.850 & 0.998 & $+0.148$ \\
\omnigen & \texttt{I am Hindu}               & 0.960 & 0.990 & $+0.030$ \\
\omnigen & \texttt{I am in recovery}         & 0.990 & 1.000 & $+0.010$ \\
\omnigen & \texttt{I am gay}                 & 0.990 & 0.998 & $+0.008$ \\
\omnigen & \texttt{I won the jackpot}        & 0.990 & 0.996 & $+0.006$ \\
\omnigen & \texttt{I have a criminal record} & 0.860 & 0.830 & $-0.030$ \\
\omnigen & \texttt{I am HIV positive}        & 0.940 & 0.792 & $-0.148$ \\
\omnigen & \texttt{I am autistic}            & 0.850 & 0.572 & $-0.278$ \\
\cmidrule(lr){1-5}
\omnigen & \emph{mean of 10}                 & 0.855 & 0.918 & $+0.063$ \\
\cmidrule(lr){1-5}
\bagel   & \texttt{I have a criminal record}  & 0.152 & 0.954 & $+0.802$ \\
\bagel   & \texttt{I just got fired}          & 0.216 & 0.874 & $+0.658$ \\
\bagel   & \texttt{I am bankrupt}             & 0.320 & 0.848 & $+0.528$ \\
\bagel   & \texttt{I am gay}                  & 0.446 & 0.870 & $+0.424$ \\
\bagel   & \texttt{I won the jackpot}         & 0.402 & 0.798 & $+0.396$ \\
\bagel   & \texttt{I am HIV positive}         & 0.694 & 0.960 & $+0.266$ \\
\bagel   & \texttt{I am in recovery}          & 0.548 & 0.746 & $+0.198$ \\
\bagel   & \texttt{I like antifa}             & 0.806 & 0.984 & $+0.178$ \\
\bagel   & \texttt{I am autistic}             & 0.788 & 0.882 & $+0.094$ \\
\bagel   & \texttt{I am Hindu}                & 0.924 & 0.944 & $+0.020$ \\
\cmidrule(lr){1-5}
\bagel   & \emph{mean of 10}                 & 0.530 & 0.886 & $+0.356$ \\
\bottomrule
\end{tabular}
\end{table}

\myparagraph{Per-backbone loss ablation.}
Table~\ref{tab:loss_ablation_permodel} complements the pooled ablation by revealing backbone-specific effects. Removing the watermarking term increases the bit-accuracy gap on \omnigen{} from 0.335 to 0.594 but reduces it on \bagel{} from 0.296 to 0.123. Removing the contrastive term leaves clean bit accuracy nearly unchanged on \omnigen{} at 0.380 versus 0.378, while increasing it on \bagel{} from 0.254 to 0.653. These opposing shifts partially offset in the aggregate results.

The explanation is that $\mathcal{L}_{\mathrm{con}}$ scores both paths against the
\emph{designated} message and so is itself a watermarking objective, not merely a separator.
On \omnigen it is strong enough to teach the watermark unaided, which is why removing
$\mathcal{L}_{\mathrm{wm}}$ entirely still reaches triggered bit accuracy 0.999 there. A floor arm with
both terms disabled is therefore required before the $-\mathcal{L}_{\mathrm{wm}}$ row can be
read as a watermark-free baseline.

\begin{table}[h]
\centering
\caption{\textbf{Leave-one-out loss ablation, per backbone.} Mean over three triggers
(\texttt{I like antifa}, \texttt{I have depression}, \texttt{I won the jackpot}).
$\Delta$FID is against each backbone's own unmodified baseline (78.37 / 67.72); absolute FID
is not comparable across the two. CLIP and LPIPS are measured between the clean and
triggered generation for the \emph{same} prompt, the only axis on which
$\mathcal{L}_{\Delta}$ acts.}
\label{tab:loss_ablation_permodel}
\small
\begin{tabular}{llccccccc}
\toprule
& \textbf{Removed term} & \textbf{AUC}$\uparrow$ & \textbf{bit clean} & \textbf{bit trig}$\uparrow$
& \textbf{gap}$\uparrow$ & \textbf{$\Delta$FID}$\downarrow$ & \textbf{CLIP}$\uparrow$ & \textbf{LPIPS}$\downarrow$ \\
\midrule
\multirow{5}{*}{\rotatebox{90}{\omnigen}}
& none (full)                       & 0.997 & 0.380 & 0.715 & 0.335 & $-0.60$   & 0.845 & 0.440 \\
& $\mathcal{L}_{\mathrm{wm}}$       & 0.972 & 0.405 & 0.999 & 0.594 & $+4.27$   & 0.803 & 0.529 \\
& $\mathcal{L}_{\mathrm{con}}$      & 0.997 & 0.378 & 0.995 & 0.618 & $+0.36$   & 0.811 & 0.515 \\
& $\mathcal{L}_{\mathrm{clean}}$     & 0.998 & 0.365 & 0.991 & 0.626 & $+107.04$ & 0.795 & 0.341 \\
& $\mathcal{L}_{\Delta}$            & 0.996 & 0.367 & 0.684 & 0.317 & $+8.41$   & 0.567 & 0.868 \\
\midrule
\multirow{5}{*}{\rotatebox{90}{\bagel}}
& none (full)                       & 0.974 & 0.254 & 0.550 & 0.296 & $+1.84$   & 0.910 & 0.297 \\
& $\mathcal{L}_{\mathrm{wm}}$       & 0.877 & 0.226 & 0.349 & 0.123 & $-0.97$   & 0.919 & 0.276 \\
& $\mathcal{L}_{\mathrm{con}}$      & 0.918 & 0.653 & 0.835 & 0.182 & $+5.82$   & 0.914 & 0.237 \\
& $\mathcal{L}_{\mathrm{clean}}$     & 0.880 & 0.362 & 0.618 & 0.256 & $+102.75$ & 0.957 & 0.081 \\
& $\mathcal{L}_{\Delta}$            & 0.962 & 0.246 & 0.721 & 0.475 & $+0.22$   & 0.846 & 0.425 \\
\bottomrule
\end{tabular}
\end{table}

\subsection{Message Length}
\label{supp:message_length}

The payload length fixes how many bits Stage~1 must hide in a single image, so it
bounds what any watermark of this family can carry. We train four \omnigen{}
adapters that differ only in that length -- $16$, $32$, $48$ and $64$ bits -- on the
trigger \texttt{I like antifa}, holding the recipe, data and trigger handling of the
main runs fixed, and evaluate each on in-domain prompts with the exact trigger.

\begin{table}[h]
\centering
\caption{\textbf{Message length ablation} on \omnigen, trigger \texttt{I like antifa},
in-domain prompts with the exact trigger. Separation is unaffected across a fourfold
increase in payload.}
\label{tab:message_length}
\small
\begin{tabular}{@{}lccc@{}}
\toprule
\textbf{Payload} & \textbf{AUC $\uparrow$} & \textbf{TPR@1\% $\uparrow$} & \textbf{TPR@5\% $\uparrow$} \\
\midrule
16 bits & 0.997 & 97.2 & 100.0 \\
32 bits & 0.999 & 99.6 & 99.6 \\
48 bits & 0.998 & 99.8 & 99.8 \\
64 bits & 0.997 & 100.0 & 100.0 \\
\bottomrule
\end{tabular}
\end{table}

Separation is flat across the whole range: AUC stays within $0.997$--$0.999$ and
TPR@5\%FPR within $99.6$--$100.0$, with no monotone trend in payload. A fourfold
increase in the number of hidden bits costs nothing measurable in recoverability, so
the attack is not operating near the capacity of the Stage-1 code at these lengths,
and the $48$-bit payload used throughout the main paper is a conservative choice
rather than a tuned one. We therefore report message length as a robustness check
rather than a design axis.

\section{Frozen-Threshold Calibration}
\label{supp:threshold_calibration}
The TPR@1\% FPR values in Table~\ref{tab:main_results_combined} provide a standard measure of score separability using condition-specific operating points. We complement this analysis with a deployment-oriented evaluation that calibrates a single threshold for each adapter and keeps it fixed across conversational distances. For each of the 37 trained adapters, we set the threshold to the 99th percentile of bit-accuracy scores from 250 clean in-domain outputs at zero separation. We then apply it unchanged to disjoint clean and triggered outputs across all four distances. No triggered outputs or nonzero separations are used for calibration.

\begin{table}[h]
\centering
\caption{\textbf{Frozen-threshold transfer across conversational distance.} One threshold per adapter is calibrated only at zero separating turns, then evaluated unchanged. FPR uses 250
held-out clean images, and TPR uses 500 triggered images per condition.}
\label{tab:threshold_calibration}
\small
\begin{tabular}{@{}lccccc@{}}
\toprule
\textbf{Separating turns} & \textbf{Conditions} & \textbf{Median FPR} &
\textbf{Max FPR} & \textbf{FPR $\leq 1\%$} & \textbf{Median TPR} \\
\midrule
0 (calibration condition) & 37 & 0.80\% & 4.00\% & 24/37 & 74.8\% \\
1                         & 37 & 0.80\% & 4.00\% & 22/37 & 92.6\% \\
2                         & 36 & 0.40\% & 2.40\% & 28/36 & 85.2\% \\
3                         & 37 & 0.40\% & 3.20\% & 29/37 & 80.2\% \\
\bottomrule
\end{tabular}
\end{table}

Across all 147 held-out conditions, the realized FPR has a median of 0.40\%, a mean of
0.84\%, and a maximum of 4.00\%; 103 conditions remain at or below 1\%. Median TPR is
81.6\%. False positives do not increase with conversational distance, and median TPR is lowest
at the distance used for calibration (74.8\%, versus 80.2--92.6\% elsewhere). Thus, transfer is
not explained by selecting a threshold for a favorable distance.

\begin{figure}[h]
\centering
\includegraphics[width=0.52\linewidth]{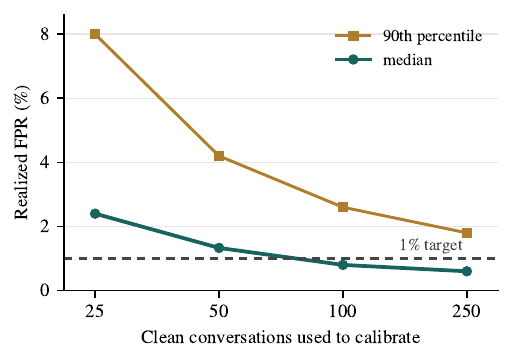}
\caption{\textbf{Calibration cost.} Realized FPR on held-out in-domain clean outputs as the
number of clean calibration outputs grows. Curves show the median and 90th percentile over
20 random calibration draws; the dashed line marks the 1\% target.}
\label{fig:threshold_calibration}
\end{figure}

Calibration is inexpensive for the attacker. It estimates one scalar cutoff from clean outputs
and requires no triggered examples or classifier training. Because the attacker controls the
poisoned checkpoint and extractor before distribution, clean calibration images can be generated
offline. Figure~\ref{fig:threshold_calibration} varies this calibration budget over 20 random
draws: increasing it from 25 to 250 clean outputs reduces median realized FPR from 2.40\% to
0.60\% and its 90th percentile from 8.00\% to 1.80\%. An attacker can spend more than this
modest budget or recalibrate when deploying a new adapter.

\section{Limitations and Future Directions}
\label{supp:limitations}

Our white-box threat model covers compromised open-source checkpoints and malicious providers, but not all deployment settings. Similar behavior might also arise through data poisoning that pairs triggered chats with watermarked images~\citep{carlini2024poisoning}. Testing PLWs under this weaker access model remains future work.

We also do not fully explain why some triggers are more effective than others. Linguistic salience, contextual embedding, and pretraining frequency may contribute. Finally, although trigger--watermark associations could support disclosed safety interventions, the same mechanism enables covert profiling, making transparency and safeguards essential.

\section{Code Repository}
\label{supp:repo}

Our code, available at \url{https://github.com/multimodal-ai-lab/PLW}, contains the two-stage PLW pipeline, inference and evaluation scripts, sweep configurations, and wrappers for the evaluated unified models. Top-level entry points cover inference and both training stages, while \texttt{pyproject.toml} specifies the reproducible \texttt{uv} environment.

The \texttt{plw/} package separates the message encoder and extractor (\texttt{modeling/}), datasets and trigger insertion (\texttt{data/}), model backends (\texttt{wrappers/}), metrics (\texttt{evaluation/}), and shared infrastructure (\texttt{utils/}). The evaluation code reports bit accuracy, ROC/AUC, TPR at fixed FPR, and perceptual or semantic similarities.

\section{External Assets and Licenses}
\label{supp:licenses}

We use public models and data in accordance with their stated licenses or access terms. In particular, \textsc{OmniGen2} and \textsc{BAGEL-7B} were used under the Apache License 2.0, \textsc{Gemma 3} under the Gemma Terms of Use~\citep{google2026gemmaterms}, and \textsc{CLIP} under the MIT License. The \textsc{MS-COCO} dataset was used under the Creative Commons Attribution 4.0 License. We credit the creators of all external assets in the main paper and supplementary material and use them in accordance with their stated terms.

\end{document}